%% file: 2020_dislocation_network_surrogate.tex
\documentclass[]{siamart1116}

\floatstyle{plaintop}
\restylefloat{table}
\input{shared}

\ifpdf
\hypersetup{
	pdftitle={\TheTitle},
	pdfauthor={\TheAuthors}
}
\fi

\begin{document}
	\maketitle
	
	\begin{abstract}
	From their birth in the manufacturing process, materials inherently contain defects that affect the mechanical behavior across multiple length and time-scales, including vacancies, dislocations, voids and cracks. Understanding, modeling, and real-time simulation of the underlying stochastic micro-structure defect evolution is therefore vital towards multi-scale coupling and propagating numerous sources of uncertainty from atomistic to eventually aging continuum mechanics. We develop a graph-based surrogate model of dislocation glide for computation of dislocation mobility. We model an edge dislocation as a random walker, jumping between neighboring nodes of a graph following a Poisson stochastic process. The network representation functions as a coarse-graining of a molecular dynamics simulation that provides dislocation trajectories for an empirical computation of jump rates. With this construction, we recover the original atomistic mobility estimates, with remarkable computational speed-up and accuracy. Furthermore, the underlying stochastic process provides the statistics of dislocation mobility associated to the original molecular dynamics simulation, allowing an efficient propagation of material parameters and uncertainties across the scales.
    \end{abstract}
% REQUIRED

\section{Introduction}

Multi-scale materials modeling and simulations are a rapidly growing scientific field, where it is critical to propagate uncertainties to accurately and efficiently bridge material properties between adjacent length- and time-scales. Among several types of material imperfections that cause disturbances in crystal structures, dislocations are line defects \cite{Hull2011} that are naturally present from manufacturing until failure of crystalline materials. Describing the small-scale buildup and dynamics of dislocations can provide an important insight on early fatigue precursors \cite{Ghoshal2019AdvancedHT,habtour2016detection}, which are beyond the resolution of existing continuum models of high-cycle fatigue damage. In order to accurately propagate such early statistics of failure to the continuum for large-scale applications, consistent, robust and efficient coupling frameworks between the atomistic and meso-scales are fundamental.

Molecular dynamics (MD) is a first-principle theory that explicitly describes the motion individual atoms at small scales based on Newton's second law. In the context of dislocations, MD has been employed as an effective tool for the atomistic understanding of canonical types of dislocation motion for diverse crystal structures and their corresponding mobility/drag coefficients \cite{Chang2001,Maresca2018,Zhou1998,Chen2020,Queyreau2011}, as well as the estimation of core energies, responsible for dislocation self-interactions \cite{lehtinen2018DDD,lehtinen2016multi}. In order to describe the complex arrangements and mechanics of dislocation networks at the intermediate scale of scanning electron microscopy \cite{Bulatov2006}, discrete dislocation dynamics (DDD) has become a practical computational tool \cite{Arsenlis2007} that allowed the discovery of new physics, such as dislocation multi-junctions \cite{Bulatov2006multijunction}. Accurate DDD simulations require precise experimental properties from dislocations and the corresponding medium, which can be obtained through MD experiments. However the large number of degrees of freedom required for robust MD simulations may render such experiments prohibitive, especially when a large number of realizations is needed to propagate the statistical qualities from small- to large-scales.

Aiming to simulate processes at longer time-scales, while still respecting the intrinsic physics of lower-scale dynamics, different approaches have emerged. Kinetic Monte Carlo (KMC) methods became popular in the last decades in a myriad of materials science applications. KMC is a type of continuous-time Markov process \cite{voter2007introduction,schulze2008efficient}, where the process rates should be known in advance. This method appeared originally for simulation of vacancies \cite{young1966monte} and Ising spin systems \cite{bortz1975new}, gaining popularity among a variety of applications, including crystal growth \cite{meng1996dynamical}, visco-elasticity \cite{baeurle2006new}, and surface kinetics \cite{andersen2019practical}. Researchers have also used KMC methods to construct low-fidelity models for dislocation motion in materials ranging from bcc metals \cite{cai2001kinetic} to Silicon \cite{cai1999kinetic,cai2000intrinsic,scarle2004linewise}, where temperature, size, and stress effects are investigated. More recently, \cite{zhao2018direct,shinzato2019atomistically} used KMC to study the interaction between solute atoms and screw dislocation bcc metals. This approach has the advantage to capture rare thermally-activated motions, which is not possible in MD simulations \cite{stukowski2015thermally}. However, such models are limited due to uncertainties in atomistic estimation of parameters used in the computation of rate constants, commonly obtained from activation energies derived from transition state theory \cite{cai2002kinetic}. Phase Field Crystal (PFC) is another fast growing method for simulation of crystalline structures with atomistic detail, while reaching diffusive time-scales, and has been used to model dislocation dynamics \cite{chan2010plasticity,asadi2015review,zaeem2018phase,ainsworth2020fractional}.

More recently, graph theory \cite{west1996introduction} has also presented itself as a robust approach in the field of materials science, with applications in coarse-graining \cite{webb2018graph}, and chemical kinetics, in combination with KMC method \cite{stamatakis2011graph}. Graph theory has a leading potential to provide efficient coarse-graining of micro-scale dynamics, furnishing suitable ground for stochastic simulations of underlying dislocation dynamics through a random walk over a network. For an extensive review of random walks on networks, we refer the reader to \cite{masuda2017random} and references therein. 

In this work, we propose a data-driven framework for the construction of a surrogate model of edge dislocation glide, where dislocation position as a time-series data is collected from high-fidelity MD simulations to train the model. We first perform a coarse-graining of the atomistic domain through a graph-theoretical formulation. In the case of dislocation glide in a periodic domain, a ring graph provides an accurate representation. However, the general construction of the network and associated operators allows further enhancements for more complex dynamics in a direct way. We model dislocation motion as a random walker, jumping between neighboring nodes on the network, following a continuous-time, Markovian stochastic process. The waiting times for forward or backward jumps between neighboring nodes is exponentially distributed with rate parameter directly computed from the MD time-series data. We supply a KMC algorithm with the estimated rate constants to simulate the dislocation motion under different applied shear stresses, providing fast and accurate calculations of dislocation velocity and mobility.

Ultimately, beyond the efficient estimates of material properties at the atomistic-level, the proposed framework allows the propagation of uncertainties across the scales. With the stochastic description of dislocation motion through a random walk over a network, governed by a Markov jump process, we can compute statistics associated to the dislocation motion that are intrinsically attached to the original atomistic setup. Mobility estimates and associated uncertainties provided by the surrogate model can later be upscaled to meso-scale dislocation simulations, such as DDD. At that stage, the collective behavior of dislocations would intrinsically incorporate stochastic effects of lower scales that would be propagated to the continuum (\textit{i.e.}, through dislocation density and plastic strains), therefore providing efficient multi-scale coupling starting in the MD domain. This feature is essential to the development of predictive models at the component level, whether the interest is on visco-elasto-plasticity \cite{suzuki2016fractional,suzuki2019thermodynamically,varghaei2019vibration}, fracture \cite{barros2020integrated,de2020data} or fatigue \cite{boldrini2016non}. 

\section{Data-Driven Framework}
\label{sec:framework}

\begin{figure}[t]
	\centering
	\includegraphics[width=\textwidth]{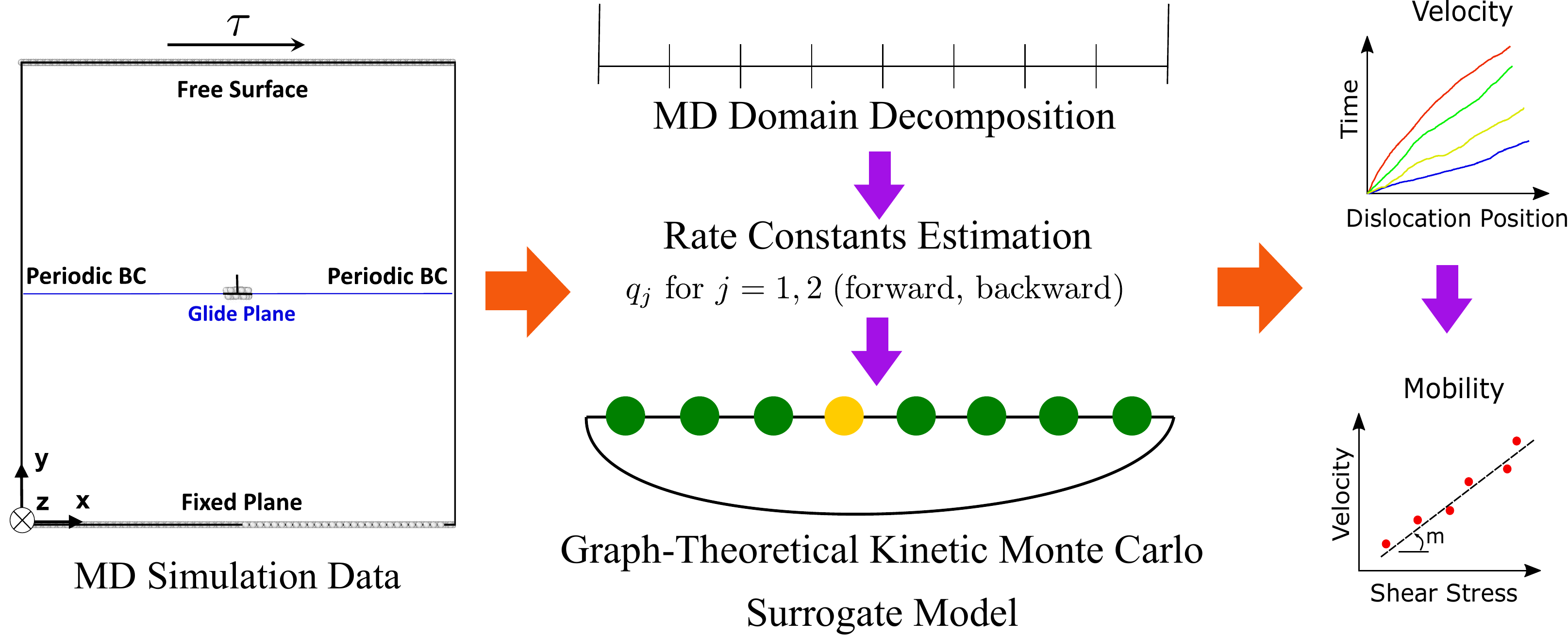}
	\caption{Framework for constructing a network-based KMC surrogate model for dislocation glide. The surrogate is then employed for fast and accurate simulations of dislocation motion, obtaining velocity data at different stress levels, leading to the estimation of the dislocation mobility.}
	\label{fig:master}
\end{figure}

We develop a surrogate model for dislocation glide parameterized by MD data to quickly obtain estimates of dislocation mobility in a short time-frame. The numerical framework for model construction and simulation is illustrated in Fig.~\ref{fig:master}. To construct the surrogate, the atomistic domain is coarse-grained and idealized as a periodic line graph (a ring graph), where nodes correspond to the sub-domains inside the crystal. 

From the coarse-grained description, we represent the dislocation as a random walker that jumps between neighboring nodes following a Poisson stochastic process. The rate constants that parameterize the process are obtained directly from MD simulation data of an edge dislocation gliding under shear stress, allowing the reconstruction and simulation of the stochastic dislocation motion through KMC method. KMC and MD are independent techniques for dislocation motion, yet here we combine both, leading to a fast computation of dislocation mobility using KMC, in which the parameters come from high-fidelity, costly, MD simulations. 

We start by discussing the methodology of dislocation simulation through MD. Then, we describe the coarse-graining of the physical domain as a ring graph, and construct the dislocation random walker based on Poisson processes. Computing the rate constants from MD simulations, we ensure that sequences of states coming from KMC converge in distribution with MD trajectories \cite{voter2007introduction}, yet using far less computation time, allowing for longer simulation times that are not achievable in MD.

\subsection{Molecular Dynamics Simulation of Edge Dislocation Glide}

Following body-centered-cubic Fe-C simulations from \cite{lehtinen2016multi}, we generate synthetic dislocation motion data in a pure Fe system and estimate the edge mobility property through MD simulations utilizing the Large-scale Atomic/Molecular Massively Parallel Simulator (LAMMPS) \cite{Plimpton1995lammps}. All the MD simulations in this work are run in 80 Intel Xeon Gold 6148 CPUs with 2.40GHz. 

\begin{figure}[t]
        \centering
        \subfloat[\label{fig:Disl_XY}]{\includegraphics[width=0.315\textwidth]{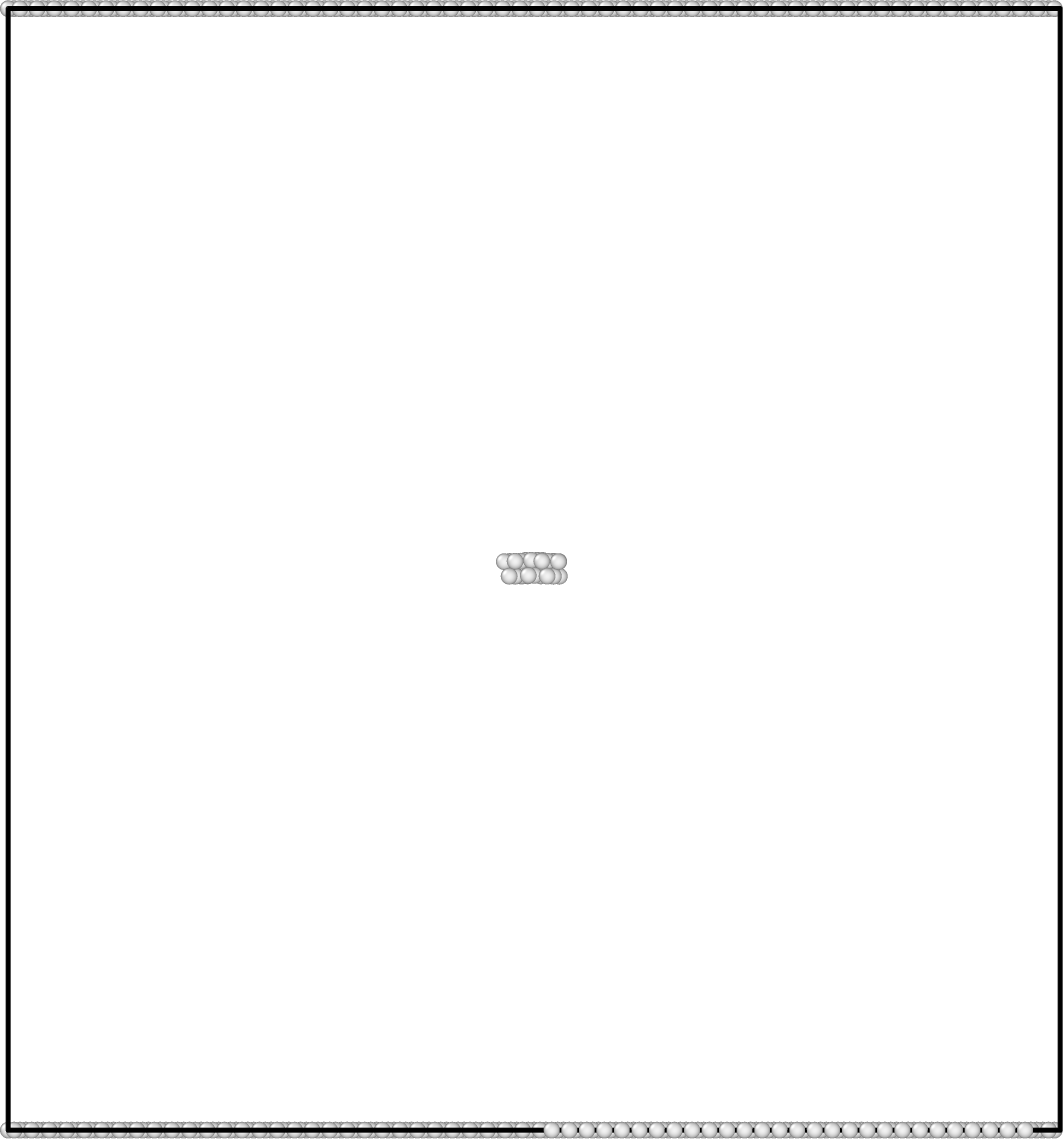}}
        \hspace{5mm}
        \subfloat[]{\includegraphics[width=0.35\textwidth]{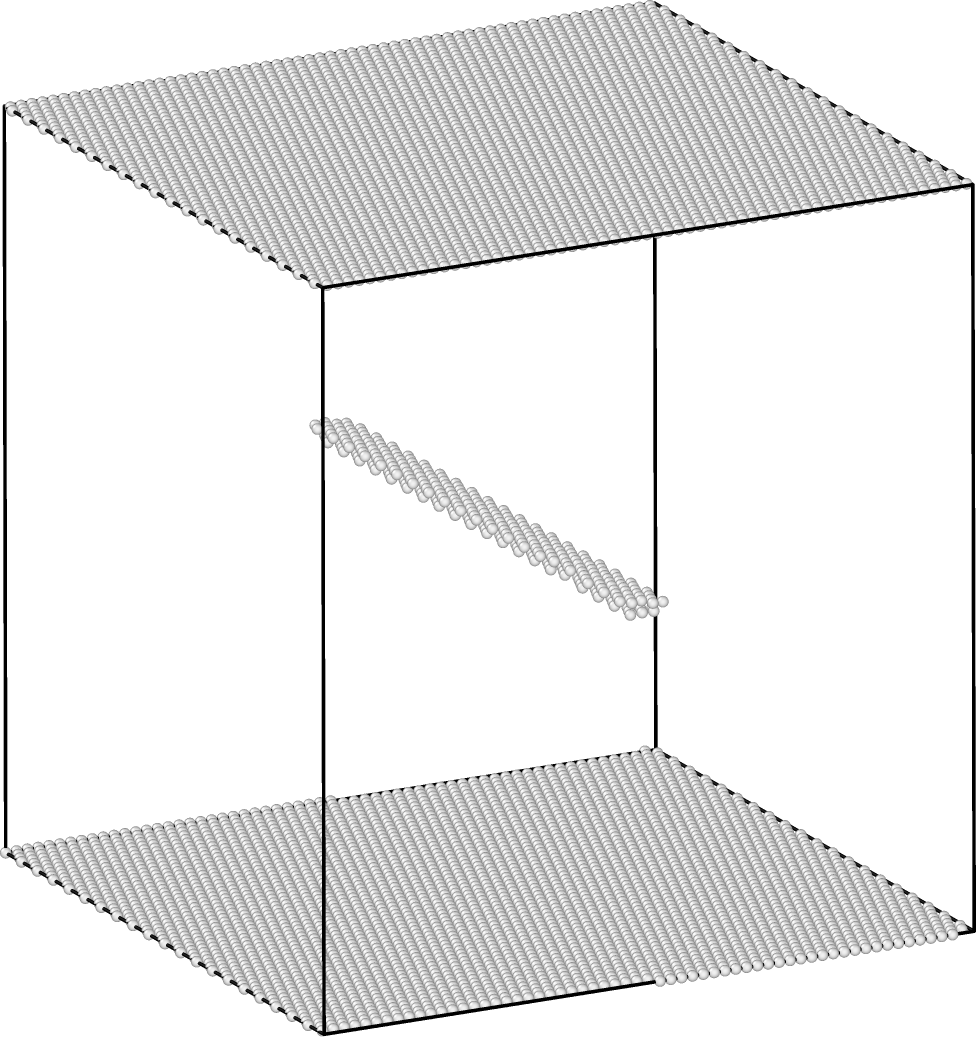}}
\caption{MD domain of the dislocation mobility test. (a) $x-y$ plane, illustrating the edge dislocation core as the lattice perturbation at the center. (b) 3D view of the MD domain with the BCC lattice removed, showing the dislocation line along the $z$-axis.}
\label{fig:Disl_glide}
\end{figure}

\begin{figure}[t]
        \centering
        \subfloat[\label{fig:temp_energy_MD}]{\includegraphics[width=0.32\textwidth]{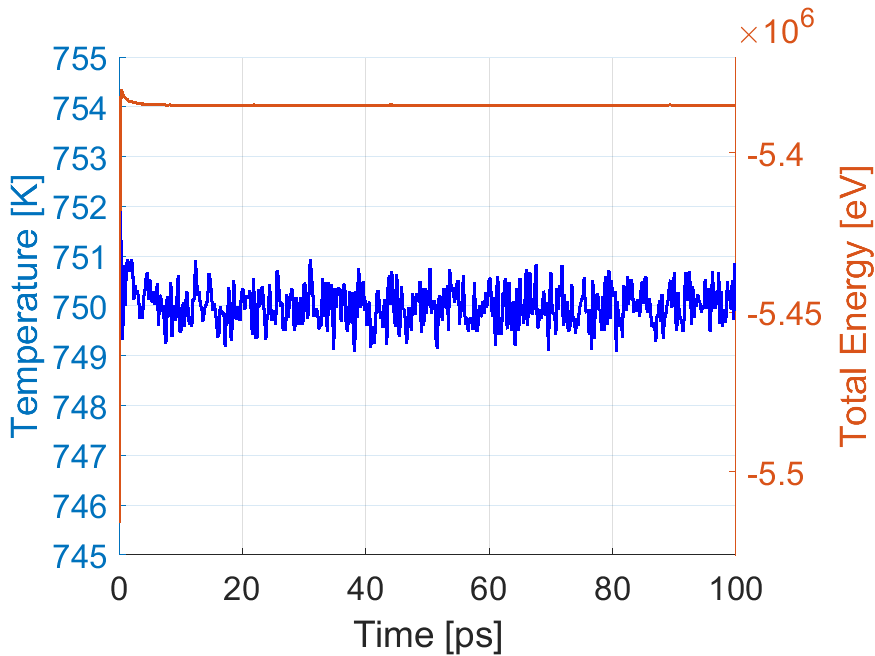}}
        ~
        \subfloat[\label{fig:pos_MD}]{\includegraphics[width=0.32\textwidth]{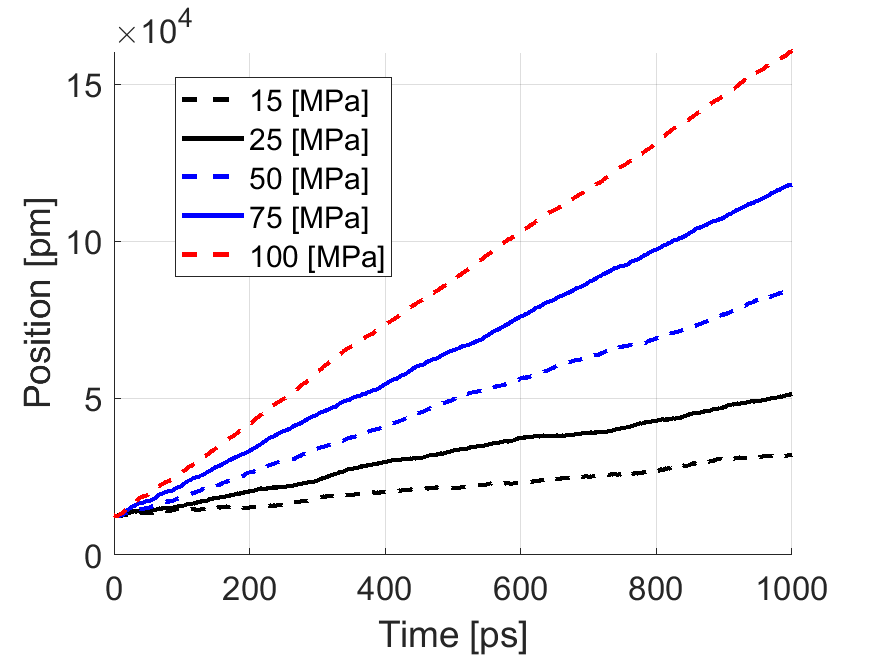}}
        ~
        \subfloat[\label{fig:mobility_MD}]{\includegraphics[width=0.32\textwidth]{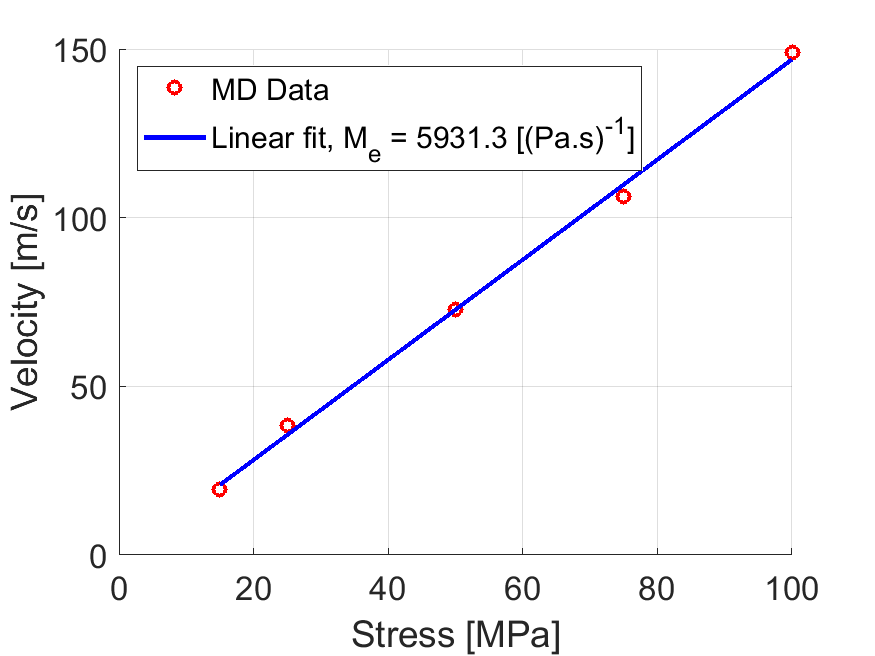}}
\caption{(a) Temperature and total energy for the equilibration step, (b) Edge dislocation position $\mathbf{x}^{MD}(t)$ and (c) mobility through MD simulations for distinct values of applied shear stresses $\tau$ under $T = 750\,[K]$. We observe an overdamped motion for the applied shear stress range and a linear mobility relationship.}
\label{fig:pos_mob_MD}
\end{figure}

The MD system under consideration is illustrated in Fig.~\ref{fig:Disl_glide}, consisting of a simulation box of $61\times 40\times 20$ $\alpha$-Fe unit cells with dimensions $25.14\times 26.96\times 24.06\,[nm]$ in the $x,\,y,\,z$ directions. A straight edge dislocation with Burgers' vector $\mathbf{b} = \frac{1}{2} [1, 1, 1]$ is generated by removing a $(1, 1, 1)$ half-plane of atoms from the center of the box. The MD domain consists of $1\,353\,132$ atoms with periodic boundary conditions applied in the $x$ and $z$ directions, and shrink-wrapped boundary conditions applied to the unit cells in the top- and bottom-planes along the $y$-direction. We perform an NVE time-integration, where the system's temperature is relaxed to $T = 750\,[K]$ through velocity-rescaling for $100\,[ps]$ \textit{(see Fig.\ref{fig:temp_energy_MD})}. We utilize a combined Tersoff bond-order and repulsive Ziegler-Biersack-Littmark (ZBL) interatomic potential, with corresponding parameters from \cite{henriksson2013atomistic}.

We apply shear stress values in the range $\tau \in [15,\,100]\, MPa$ to the top layer in Fig.~\ref{fig:Disl_XY}, parallel to $\mathbf{b}$, which induces a glide motion in the $x$-direction on the $(1, 1, 0)$ plane. No temperature control is enforced in this stage and we run the simulation over $1\,[ns]$ with time-step size $\Delta t^{MD} = 2\,[fs]$. The MD time-series data is saved every 100 time-steps and the atom positions are post-processed utilizing the Polyhedral Template Matching (PTM) method \cite{Larsen2016} implemented in OVITO (\url{https://www.ovito.org/}) \cite{Stukowski2010}, which allows us to detect and track the lattice disturbance. We define the dislocation position as the average of all $x$-coordinates of atoms belonging to the disturbed region (dislocation core) in Fig.\ref{fig:Disl_XY}. Therefore, for every applied shear stress $\tau$, we obtain a position vector $\mathbf{x}^{MD}(t)$ with 5000 data-points (\textit{see Fig.~\ref{fig:pos_MD}}) of size $\Delta t^{MD} = 2\,[fs]$, from which we compute the corresponding velocity $v^{MD}$ through a linear fit. The obtained velocity from the post-processed MD simulation can be related to the one-dimensional solution from dislocation dynamics denoted by $v_x$, and given by the following relationship:
\begin{equation}
	v_x = M \cdot b \cdot \tau.
	\label{eq:mobility_rule}
\end{equation}
where $M$ denotes the edge dislocation mobility, and $b = \sqrt{3} a/2$ represents the magnitude of $\mathbf{b}$. Equation (\ref{eq:mobility_rule}) is obtained from the balance between the applied Peach-Kohler force induced by the shear stress $\tau$ and the dislocation drag force. Therefore, setting $v^{MD} = v_x$ and from the slope $m = M \cdot b$ of the velocity \textit{versus} stress curve in Fig.~\ref{fig:mobility_MD}, we estimate the edge dislocation glide mobility as $M = m/b \approx 5931.3\,[(Pa.s)^{-1}]$, which is in good quantitative agreement (1.73\% difference) compared to the results obtained by Lehtinen et al.\cite{lehtinen2016multi}.

\subsection{Graph-theoretical Coarse-graining}

We begin the surrogate framework by idealizing a coarse-grained version of the atomistic domain as a graph $G(V,E)$, with a set vertices (or nodes) $V$ connected by edges $E$. In this representation, each node on the network represents a sub-domain from the original MD system. In the case of a dislocation glide along a single slip plane, a one-dimensional ring graph is an adequate simplification of the dislocation movement, also assuring the periodicity presented in the MD domain.

The coarse-graining is achieved by dividing the domain into $n$ sub-domains, or bins, such that

\begin{equation}
n = \left \lfloor \frac{L}{\max(\bm{d})} \right \rfloor
\label{eq:nodes}
\end{equation}

\noindent where $L$ is the size of the domain (in the $x-y$ plane), and $\bm{d}$ is the vector containing the distance traveled by the dislocation between each MD time-step, with entries $d_i = x^{MD}_{i+1} - x^{MD}_i$. We choose this upper bound to ensure that the dislocation only travels to neighboring nodes. In this sense, we identify the dislocation as corresponding to node $i$ of the graph if the dislocation position in the MD simulation lies between the bounds of bin $i$ of width $\Delta x = L/n$.

We now define the standard operators for a continuous-time random walk on a network. The adjacency matrix $\bm{A}$ has elements $A_{ij} = 1$ if there is a link between nodes $i$ and $j$, and $A_{ij} = 0$ otherwise, for $i,j = 1,\,2,\,\dots,\,N$. The degree matrix $\bm{K}$ represents the number of edges attached to the node, computed as $K_{ii} = \sum_{j=1}^N A_{ij}$, and $K_{ij} = 0$ for $i\neq j$. From $\bm{A}$ and $\bm{K}$ we define the transition matrix $\bm{W}$, with elements $w_{i\to j} = \frac{A_{ij}}{K_i}$, representing the probability of the random walker to transition from node $i$ to node $j$. 

Specifically for the ring graph considered for the surrogate model, every node is attached to two other nodes, which makes entries $A_{ii+1} = A_{ii-1} = 1$, except when $i = 1$ or $i = N$. In those cases, $A_{1n}$ and $A_{n1}$ are set to one to ensure periodicity. As a consequence, degree matrix $\bm{K}$ has all entries $K_{ii} = 2$.

The transition matrix is finally computed with elements $w_{i\to i} = w_{i\to i+1} = w_{i\to i-1} = \frac{1}{2}$. Again, the exception is for nodes $i=1$ and $i=N$, where we obtain $w_{1\to N} = \frac{1}{2}$ and $w_{N\to 1} = \frac{1}{2}$, respectively, due to periodicity. At this point we use the transition matrix $\bm{W}$ as a building block for introducing the dynamics of dislocation motion. Its purpose is to initially restrict the movement of the random walker to the neighboring nodes with equal probability, later modulated by empirical rates computed from MD simulations.

\subsection{Construction of the Random Walk}

The construction of the random walk representative of dislocation glide is dependent on two main aspects: first, on the simplification of dislocation motion and its coarse-graining through a graph-theoretical framework, as discussed before; second, on the statistical representation of dislocation mobility through a Poisson process that naturally leads to the use of KMC method. We now discuss the formulation of the random walk, where we will follow closely the ideas in \cite{voter2007introduction}.

The main attractiveness of KMC is the simplification of complex dynamics into a counting process, where the entire system moves from state to state. For each possible escape path from the current state, there is an associated rate constant $q_{ij}$, which is the probability per unit time to transition to state $j$ from state $i$. 

For modeling the dislocation motion through a random walk, we first assume that the transition probabilities for dislocation motion are independent of history, therefore, characteristic of a Markov processes. Second, for systems such as the pure Fe-Fe studied in this work, there is no evident acceleration of dislocation in the long range. Therefore, we assume that the underlying process is stationary with independent increments.

Let $(\Omega,\mathcal{F},\mathbb{P})$ be a complete probability space, where $\Omega$ is the space of outcomes $\omega$, $\mathcal{F}$ is the $\sigma$-algebra
and $\mathbb{P}$ is a probability measure, $\mathbb{P} : \mathcal{F} \to [0; 1]$. From the assumptions, we model the total number of jumps $N_t(t)$ between states over time $t\in[0,\infty)$ as a Poisson process with total rate $Q$, such that for any $t$, $N_t(t) \sim \text{Poisson}(Qt)$. 

For an arbitrary process with several possible states $j$ from current state $i$, each with rate $q_{ij}$, the total rate $Q$ is $Q = \sum_j q_{ij}$, following the assumption that the different processes are independent and non-overlapping. In the dislocation motion studied here, there are only two possible escape paths from any current state, a forward or backward jump, with respective rates $q_f$ and $q_f$. Therefore, we have

\begin{equation}
    Q = q_f + q_b.
    \label{eq:rates}
\end{equation}

Furthermore, let $X: \Omega \to \mathbb{R}$ be a random variable that represents the waiting times between jumps over the graph $G$. Then, $X \sim \text{Exponential}(Q)$, meaning that the process is first-order with exponential decay statistics, i.e., memoryless. The probability of the random walker not performing any jump, therefore staying on the current node, is given by

\begin{equation}
p_{\text{stay}} (t) = e^{-Q t},
\label{eq:p_stay}
\end{equation}

\noindent leading to the standard computation of time increments $\Delta t$ in KMC algorithms,

\begin{equation}
\Delta t = -\frac{\ln(r)}{Q},
\label{eq:time}
\end{equation}

\noindent where $r$ is a random number sampled from the uniform distribution $\mathcal{U}(0,1)$.

After each time-step with size given by Eq.(\ref{eq:time}), the system will evolve to a new state with probability proportional to $q_f$ and $q_b$. In general, this is accomplished by recomputing the elements of $\bm{W}$ as $p_{ij}$, representing the probability of a jump per unit of time, in units of $s^{-1}$. Probabilities are obtained through

\begin{equation}
p_{ij} = \frac{w_{i\to j} q_j}{\sum_j w_{i \to j}q_j} ,
\label{eq:prob_rate}
\end{equation}

\noindent where $p_{ij}$ is now the walker's probability to go from node $i$ to node $j$, per unit time . The result is normalized to make $\sum_j p_{ij} = 1$. Equivalently, we may simply take

\begin{equation}
p_{ij} = \frac{q_j}{\sum_j q_j} = \frac{q_j}{Q}
\label{eq:prob_rate_2}
\end{equation}

\noindent for $q_j \in \{q_f,q_b\}$.

\begin{remark}
    Note that the increment in time and the selection of the next state are independent of each other. First the system waits for any jump with probability related to the total jump rate $Q$. Then, in a separate drawing, the next state is chosen with probabilities proportional to $q_f$ and $q_b$.
\end{remark}

\begin{remark}
	The general graph-theoretical description of the physical system allows flexibility and future incorporation of more complex cases, beyond the ring graph currently adopted for the case of dislocation glide. The inclusion of inhomogeneous Poisson processes (either in time or space), dislocation climb, or even non-Markovian network dynamics as in the case of L\'evy flights \cite{riascos2014fractional} can be built on top of this fundamental framework in a straight-forward fashion.
\end{remark}

Since the graph nodes are positioned in the center of each bin, as illustrated in Fig.~\ref{fig:master}, we have an approximation for distance traveled by the dislocation from the internodal distance $\Delta x$. Then, at each time-step, the dislocation spatial position is updated by 

\begin{equation}
\begin{cases}
x^{n+1} = x^n + \Delta x,\quad \text{if dislocation jumps forward}\\
x^{n+1} = x^n - \Delta x,\quad \text{if dislocation jumps backwards}
\end{cases}
\label{eq:jump_update}
\end{equation}

\noindent where $x^{n+1}$ is the new dislocation position ate time-step $t_{n+1}$. In that sense, this model is still a discrete-space random walk, which calls for extra care when computing the dislocation velocity.

One possibility is to mimic the procedure from MD simulations, and plot the dislocation distance as a function of time, performing a linear regression to obtain the dislocation velocity $v$. We run the simulation for each stress level, and plot the dislocation velocity as a function of stress. Again, a linear regression is used to obtain the slope of the curve $m$ for a linear mobility rule as in MD, and the dislocation mobility from the network dynamics is estimated through Eq.(\ref{eq:mobility_rule}).

Algorithm summarizes the procedure of running a KMC simulation of dislocation glide through a random walk on a network for a total of $M$ time-steps when we know the rates of forward $q_f$ and backward $q_b$ jumps.

\begin{algorithm}[t]
	\caption{Kinetic Monte Carlo method for Dislocation Glide as a Random Walk on a Graph}
	\label{algo:kmc}
	\begin{algorithmic}[1]
		\STATE Given: rates for jump forward $q_f$ and jump backward $q_b$, compute total rate through Eq.(\ref{eq:rates}).
		\STATE Given: number of nodes $n$ through Eq.(\ref{eq:nodes}), and the distance between nodes $\Delta x$, compute transition matrix $W$. 
		\FOR{Time-steps $m = 0 \to M-1$}
		\STATE Given the current node position $i$, get the corresponding $i-$th line of $\bm{W}$.
		\STATE Update line $\bm{W}_i$ as in Eq.(\ref{eq:prob_rate_2}).
		\STATE Choose next position $j$ based on the pdf given by $W_i$.
		\STATE Generate a random number $r \sim \mathcal{U}(0,1)$.
		\STATE Advance time by a time-step $\Delta t$ from Eq.(\ref{eq:time}).
		\STATE Update the dislocation's spatial position by $\Delta x$ using Eq.(\ref{eq:jump_update}).
		\ENDFOR
	\end{algorithmic}
\end{algorithm}

\subsection{Empirical Computation of Rate Constants}

One of the major drawbacks of KMC methods is the required knowledge of process rates as inputs to the method, which may not always be a trivial task, where traditional approaches involve the computation of rates through physical principles \cite{cai2002kinetic,voter2007introduction}. In this work, we propose a data-driven approach for the computation of jump rates from dislocation position data obtained in MD simulations. In this way, the atomistic, high-fidelity simulation  with observable dislocation motion parameterizes the surrogate model through the rate constants.

From the coarse-graining procedure, at each time-step we can identify and track the node associated with the dislocation position in MD. With this information, we are able to compute the waiting times between two consecutive jumps, classified in three main groups of events: forward, backward, or any jump. We also compute the total number of jump events in any of the three groups, respectively $N_f$, $N_b$, and $N_t = N_f + N_b$. Both groups of data can be used to estimate the rate constants. 

We model $N_t(t)$ following a Poisson distribution, and given that the expectation of a Poisson random variable with parameter $\lambda = Qt$ \cite{grimmett2014probability} is 

\begin{equation}
\mathbb{E}[N_t(t)] = Q t,
\end{equation}

\noindent we may infer the rate parameter $Q$ from empirical data by taking 

\begin{equation}
Q = \frac{\mathbb{E}[N_t(t)]}{t}.
\label{eq:rate}
\end{equation}

The expected number of jumps $\mathbb{E}[N_t(t)]$ is taken here to be the number of jumps that occurred in the MD simulation during simulation time $t$. Equivalently, we can replace $N_t(t)$ by $N_f$ and $N_b$, to respectively compute  $q_f$ and $q_b$.

Alternatively, we can look at the probability that a jump happened by time $t'$, which is the integral of the probability of the first jump $p(t)$, and it is given by

\begin{equation}
\int_0^{t'} p(t) dt = 1 - p_{\text{stay}} (t').
\end{equation}

It follows that $p(t)$ can be obtained by taking $p(t) = -\frac{\partial p_{\text{stay}} (t)}{\partial t}$, so that

\begin{equation}
p(t) = Q e^{-Q t},
\label{eq:p_of_t}
\end{equation}

\noindent which is an exponential distribution of waiting times. Taking the first moment of Eq.(\ref{eq:p_of_t}) gives the average waiting time for a jump $\mu$ as

\begin{equation}
\mu = \int_0^\infty t p(t) dt = \frac{1}{Q}.
\label{eq:tau}
\end{equation}

Note that again we may generalize the result from Eq.(\ref{eq:tau}) to average waiting time between two consecutive forward and backward jumps exclusively, $\mu_f$ and $\mu_b$, just by isolating those events from the complete time-series of waiting times. In that case, we can also obtain $q_f$ and $q_b$ from waiting time distributions.

The last method we may use to compute the rate constants is also through distributions of waiting times. Yet, this time we fit an exponential function to the histogram of waiting times using Maximum Likelihood Estimation (MLE). The MLE estimator for an exponential fit is equivalent to the reciprocal of sample mean, i.e. $1/\mu$, therefore we can expect identical results when using both methods \cite{evans2004probability}. We compare the accuracy of all three methods in the following section by using user-defined true rates as reference solution.

\section{Results and Discussion}
\label{sec:results}

We now present the numerical results from the surrogate model simulations. We start by investigating the accuracy of the rate estimation algorithm, and the convergence as a function of the number of time-steps from the original data-set using manufactured known process rates. Then, we apply the framework to real MD simulation data of dislocation glide and compute the mobility using the surrogate, comparing the results with MD mobility computations.

\subsection{Convergence of Rate Constant Estimation}

We investigate the accuracy and convergence of the rate estimation algorithm through KCM simulation of a single random walker in a ring graph, with manufactured true rates $q_{\text{true}}$ for forward and backward jumps. We test different rate combinations for the jumps, and apply Eqs.(\ref{eq:rate}) and (\ref{eq:tau}), and MLE to estimate the original rates in one realization of the stochastic process. We check the convergence of the rate estimate with different number of time-steps, which in this case is the exact number of total jumps $N_t(t)$. We consider a graph with $n = 20$ nodes.

We show results in Tables~\ref{tab:2rates_200_1} and \ref{tab:2rates_100_100}, for the estimation through Eq.(\ref{eq:rate}). The other two methods yield identical results for the manufactured solution, and are omitted. We present the estimated rates $q_{\text{est}}$, and the relative error to the true rates, computed as

\begin{equation}
\text{error} = \frac{\vert q_{\text{true}} - q_{\text{est}}\vert}{\vert q_{\text{true}}\vert}.
\end{equation}

We observe that accuracy is dependent on the number of time-steps, which is natural, since more time-steps provide more data for a reliable statistical representation of the true rates. Second, the estimate is more accurate for higher rates, relative to lower ones, as in Table~\ref{tab:2rates_200_1}, where the ratio between the rates is large. For rates of similar magnitude, error levels are comparable, since there is sufficient data for both estimates.

\begin{table}[t]
	\centering
	\caption{True rates: 200 (forward) and 1 (backward), in units of $s^{-1}$.}
	\label{tab:2rates_200_1}
	\begin{tabular}{l|l|l|l|l}
		\hline Number of time-steps & Forward Rate & Error   & Backward Rate & Error   \\ \hline
		$10^1$                 & 191.9520     & 4.02\% & 0.0000        & 100.00\% \\
		$10^2$                 & 202.8340     & 1.42\% & 0.0000        & 100.00\% \\
		$10^3$                 & 214.0373     & 7.02\% & 1.9438        & 94.38\%  \\
		$10^4$                 & 197.1265     & 1.44\% & 0.9309        & 6.91\%   \\
		$10^5$                 & 199.2291     & 0.39\% & 0.9066        & 9.34\%   \\
		$10^6$                 & 200.0162     & 0.01\% & 0.9272        & 7.28\% 
	\end{tabular}
\end{table}

\begin{table}[t]
	\centering
	\caption{True rates: 100 (forward) and 100 (backward), in units of $s^{-1}$.}
	\label{tab:2rates_100_100}
	\begin{tabular}{l|l|l|l|l}
		\hline Number of time-steps & Forward Rate & Error   & Backward Rate & Error   \\ \hline
		$10^1$                 & 51.5011      & 48.50\% & 51.5011       & 48.50\% \\
		$10^2$                 & 124.4437     & 1.82\%  & 101.8176      & 1.82\%  \\
		$10^3$                 & 108.6405     & 3.66\%  & 96.3416       & 3.66\%  \\
		$10^4$                 & 98.6444      & 0.40\%  & 100.3960      & 0.40\%  \\
		$10^5$                 & 100.3814     & 0.32\%  & 99.6812       & 0.32\%  \\
		$10^6$                 & 99.8083      & 0.04\%  & 100.0381      & 0.04\% 
	\end{tabular}
\end{table}

\subsubsection{Uncertainty quantification of rate estimation}

Due to the probabilistic nature of this framework, results from Tables~\ref{tab:2rates_200_1} and \ref{tab:2rates_100_100} show oscillations in error measures, which only represent the accuracy of a single realization of the problem in the stochastic space. This motivates an Uncertainty Quantification (UQ) analysis, where we using Monte Carlo method to quantify the level of uncertainty in the rate estimation for data with different number of time-steps. 

\begin{figure}[t]
	\centering
	\subfloat[Number of realizations.]{\includegraphics[width=0.5\textwidth]{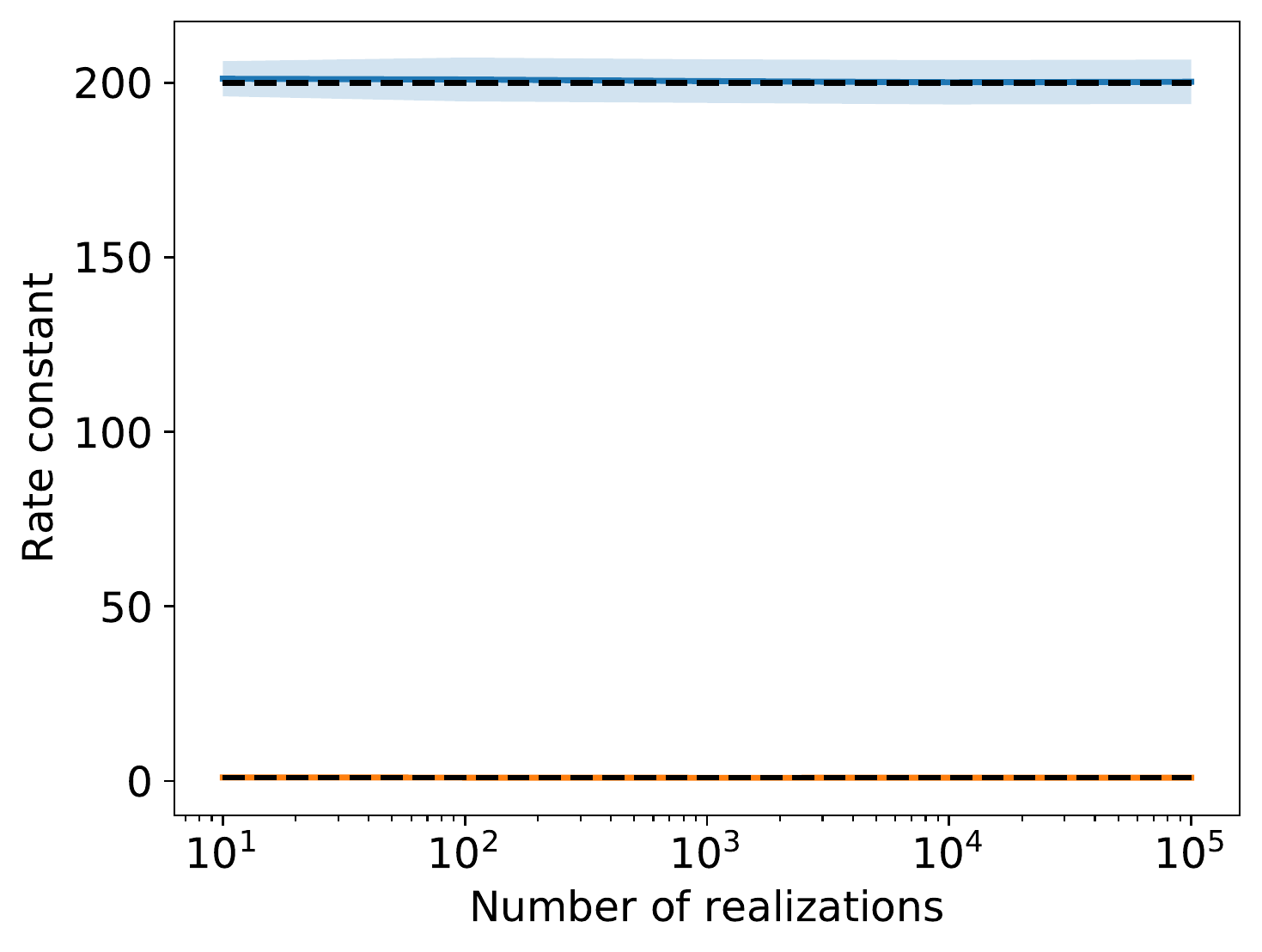}}
	\subfloat[Number of time-steps.]{\includegraphics[width=0.5\textwidth]{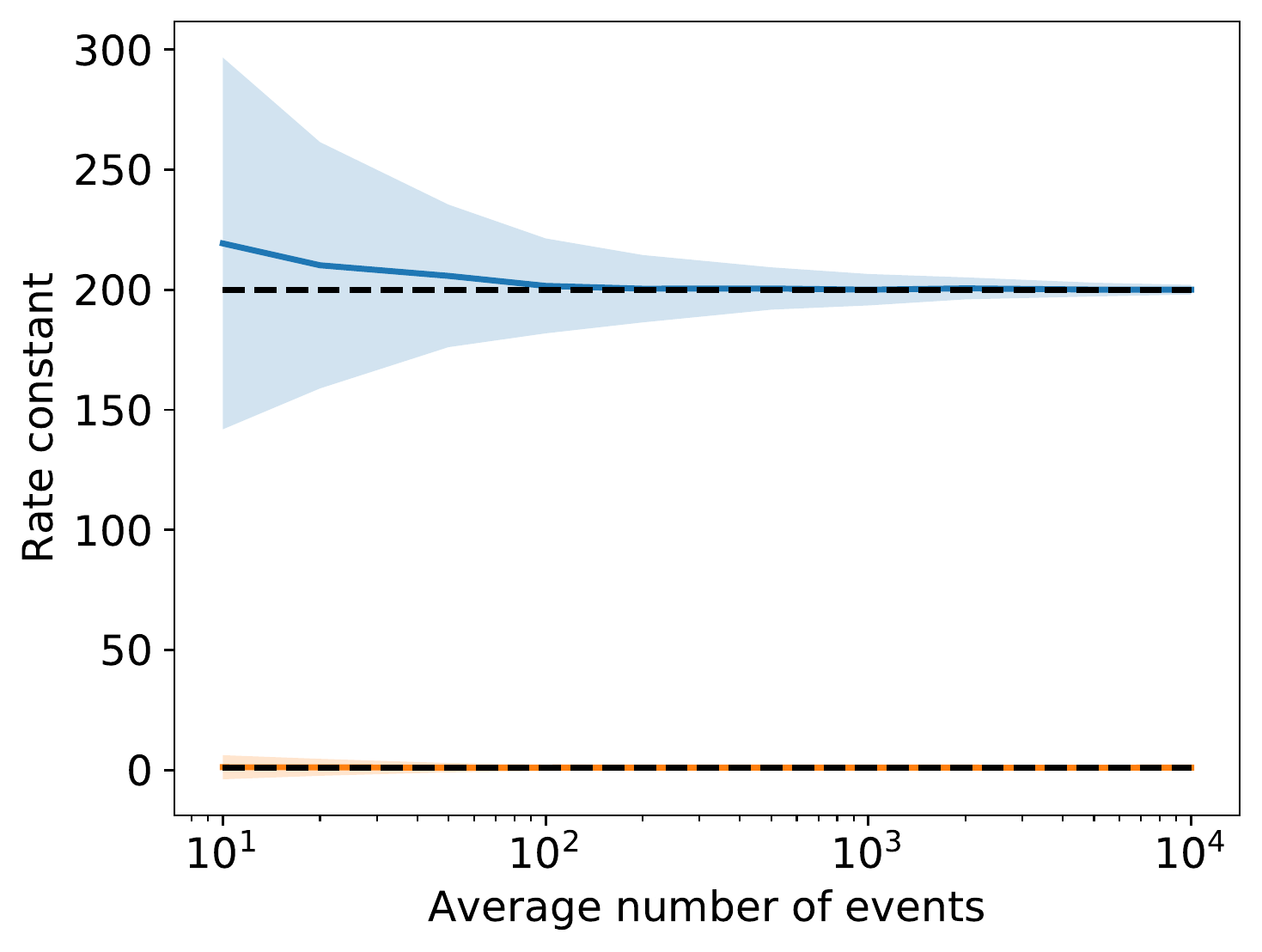}}
	\caption{Convergence to true rates (y-axis) as a function of number of realizations with fixed time-steps (a), or number of time-steps with fixed realizations (b). Dashed lines are the true rates (200 and 1), solid lines are the expected rates, and the shaded areas are the regions of uncertainty based on standard deviation.}
	\label{fig:mc}
\end{figure}

Two types of analysis were performed. First, for a fixed number of 1000 time-steps, expectation and standard deviation were obtained for different number of MC realizations. Last, for a fixed number of 1000 realizations, we obtained expectation and standard deviation for different number of time-steps, i.e., by considering different final simulation times from the time-series data. To show this result, we average the number of statistical events (total jumps) from each realization to construct the $x$-axis. For the same network of $n = 20$ nodes as before, and true rates of $q_{f,\text{true}} = 200\ s^{-1}$ and $q_{b,\text{true}} = 1\ s^{-1}$ as in Table~\ref{tab:2rates_200_1}, we plot UQ results in Fig.~\ref{fig:mc}.

From Fig.~\ref{fig:mc} we see that the precise computation of rates from data is almost exclusively dependent on the number of statistical events, therefore on the length of the simulation. Increasing the number of realizations does not increase the accuracy of the recovered rates, and the uncertainty region is kept constant. However, increasing the number of time-steps through considering longer simulation times leads to the expected value to converge to the true rate, and shrinks the uncertainty region.

\subsection{Dislocation Mobility}

Here we present numerical results for one complete cycle of the framework, from MD simulation of dislocation glide, to rate estimation and final surrogate simulation through a random walk in the constructed network.  

\subsubsection{Rate estimation}

From the raw data of dislocation position and time obtained from LAMMPS and Ovito, we apply the domain decomposition into bins equivalently to graph nodes, and track the current node over time. We count the number of jumps forward and backward between two nodes, as well as the waiting times between events. This also allow us to compute the waiting times between two forward or backward jumps. 

\begin{figure}[t]
	\centering
	\subfloat[]{\includegraphics[width=0.33\textwidth]{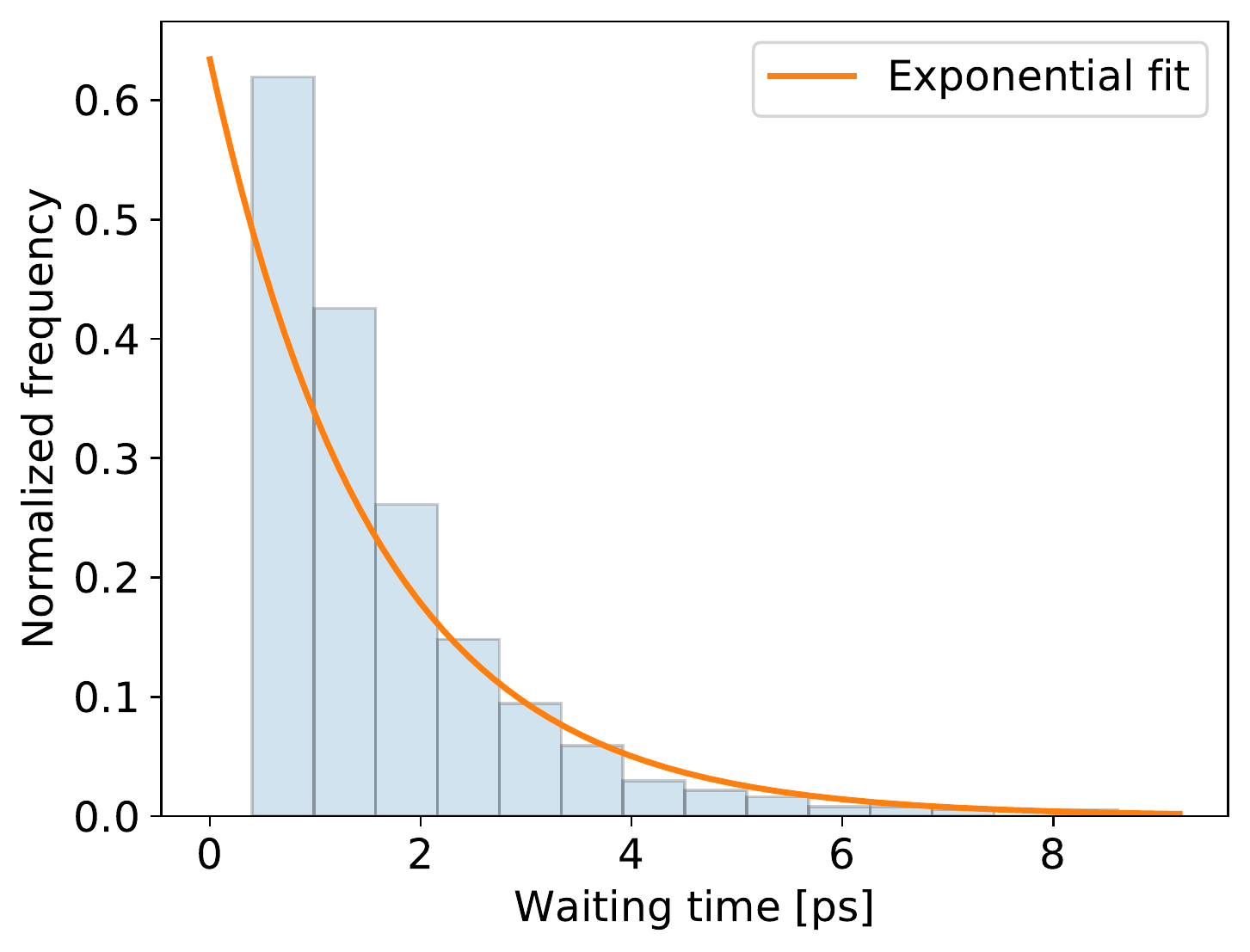}}
	\subfloat[]{\includegraphics[width=0.33\textwidth]{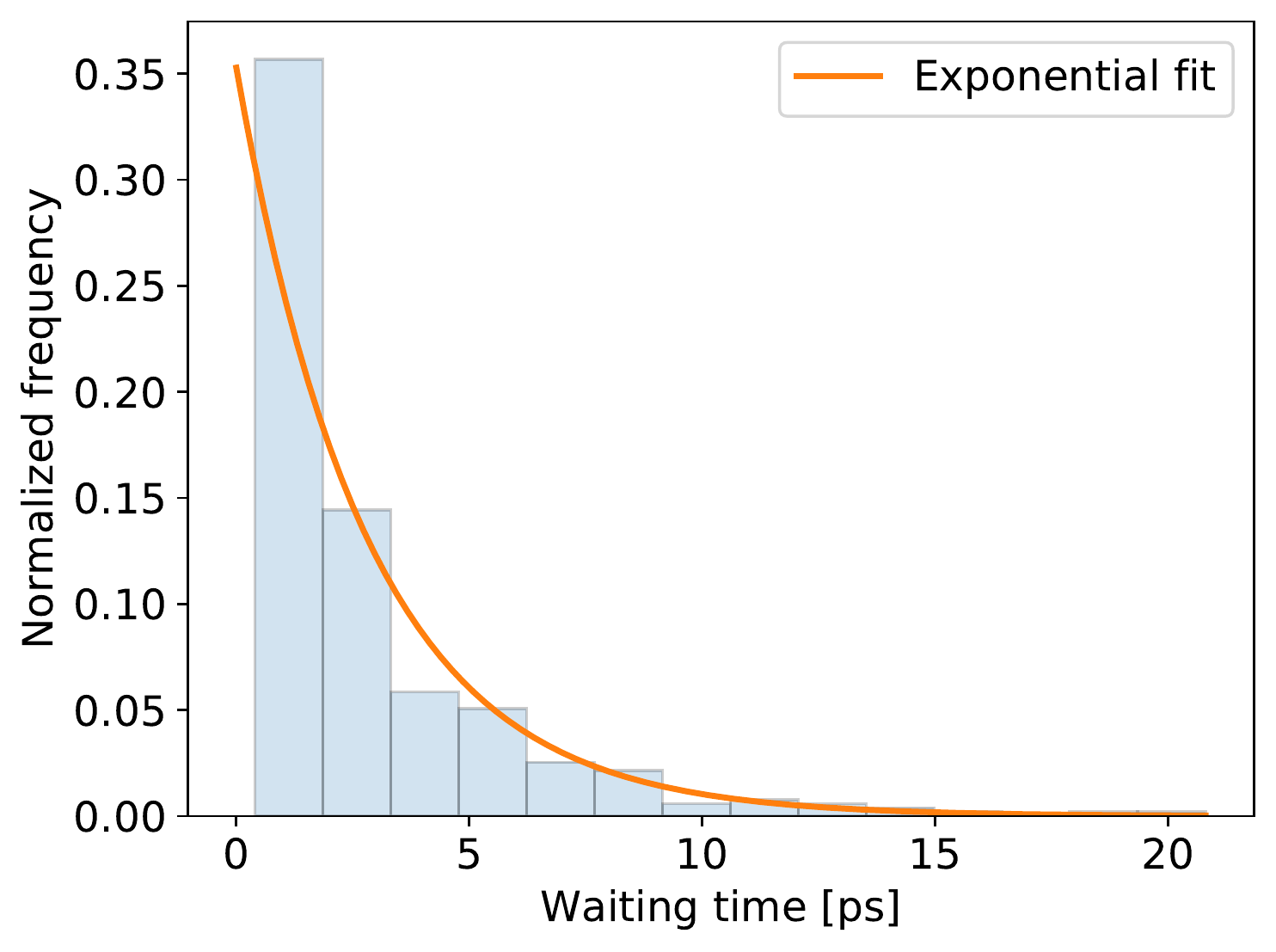}}
	\subfloat[]{\includegraphics[width=0.33\textwidth]{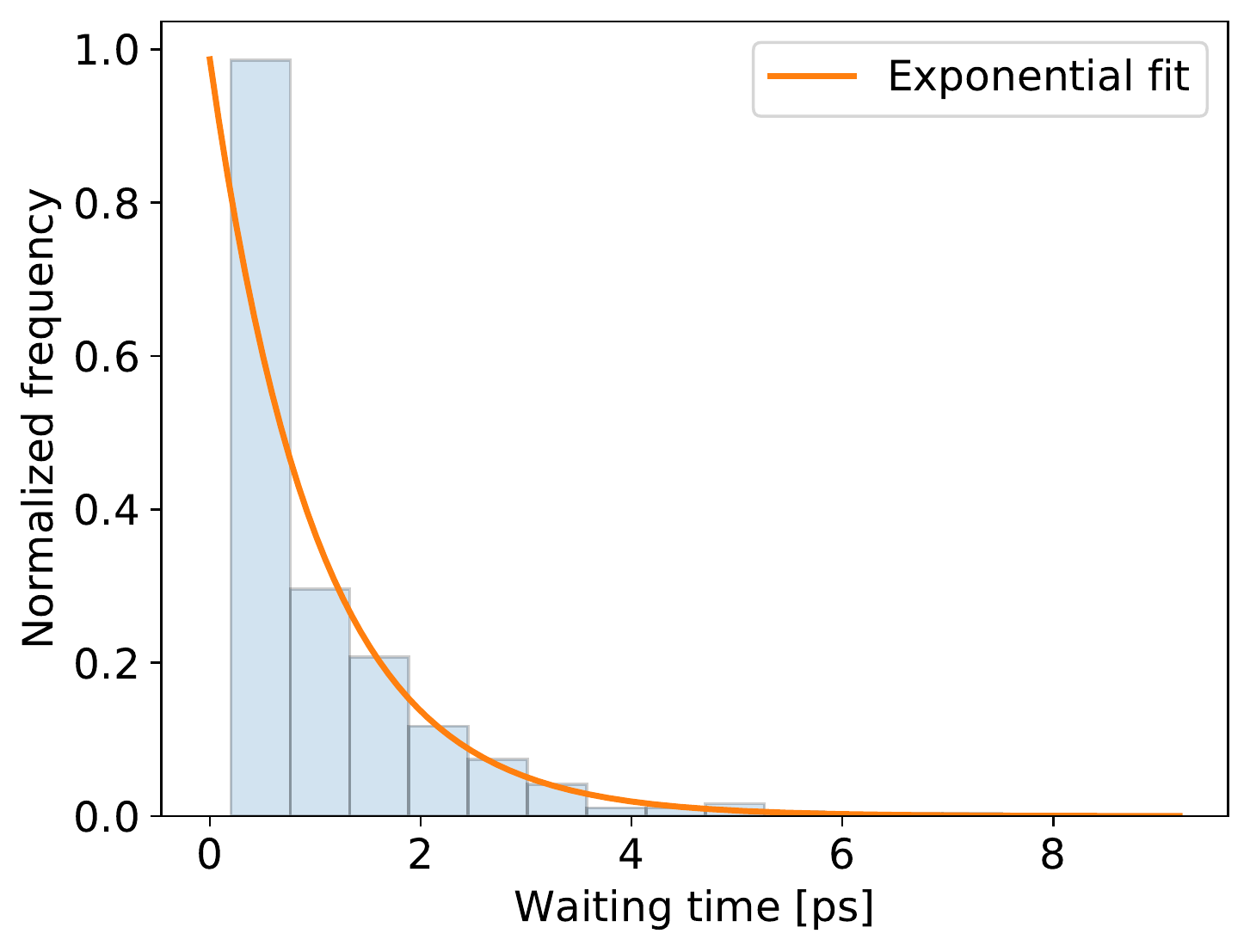}}\\
	\subfloat[]{\includegraphics[width=0.33\textwidth]{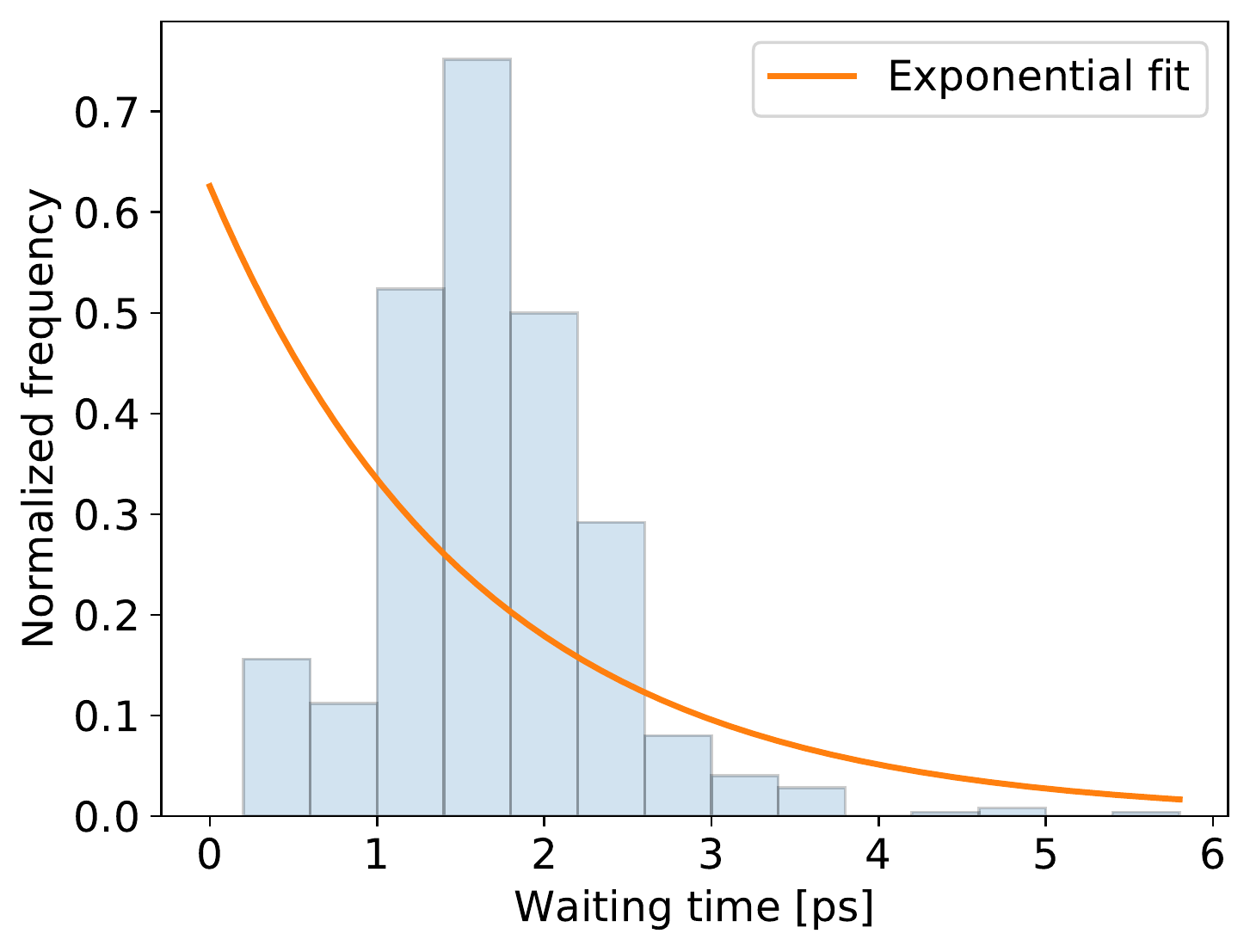}}
	\subfloat[]{\includegraphics[width=0.33\textwidth]{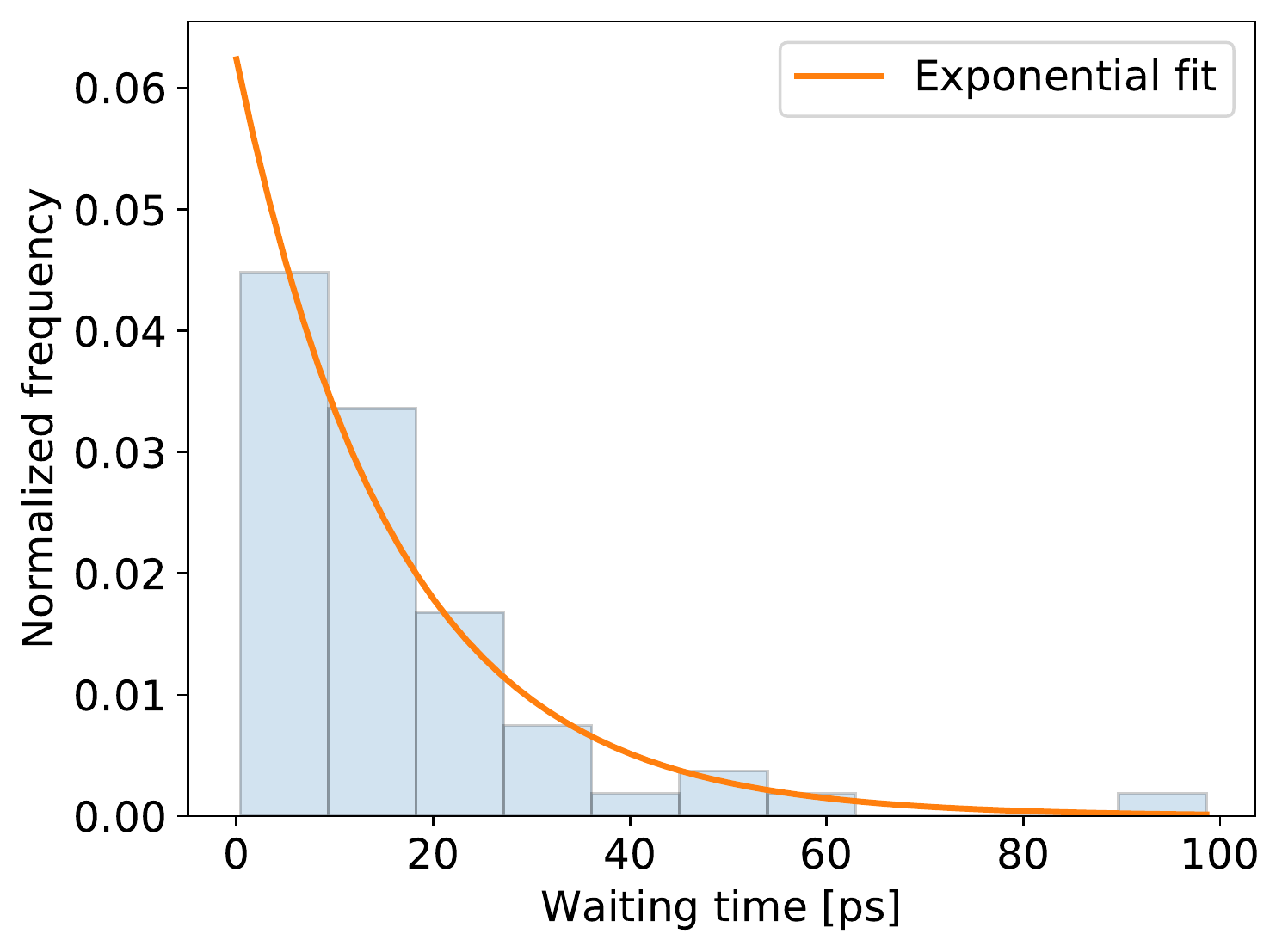}}
	\subfloat[]{\includegraphics[width=0.33\textwidth]{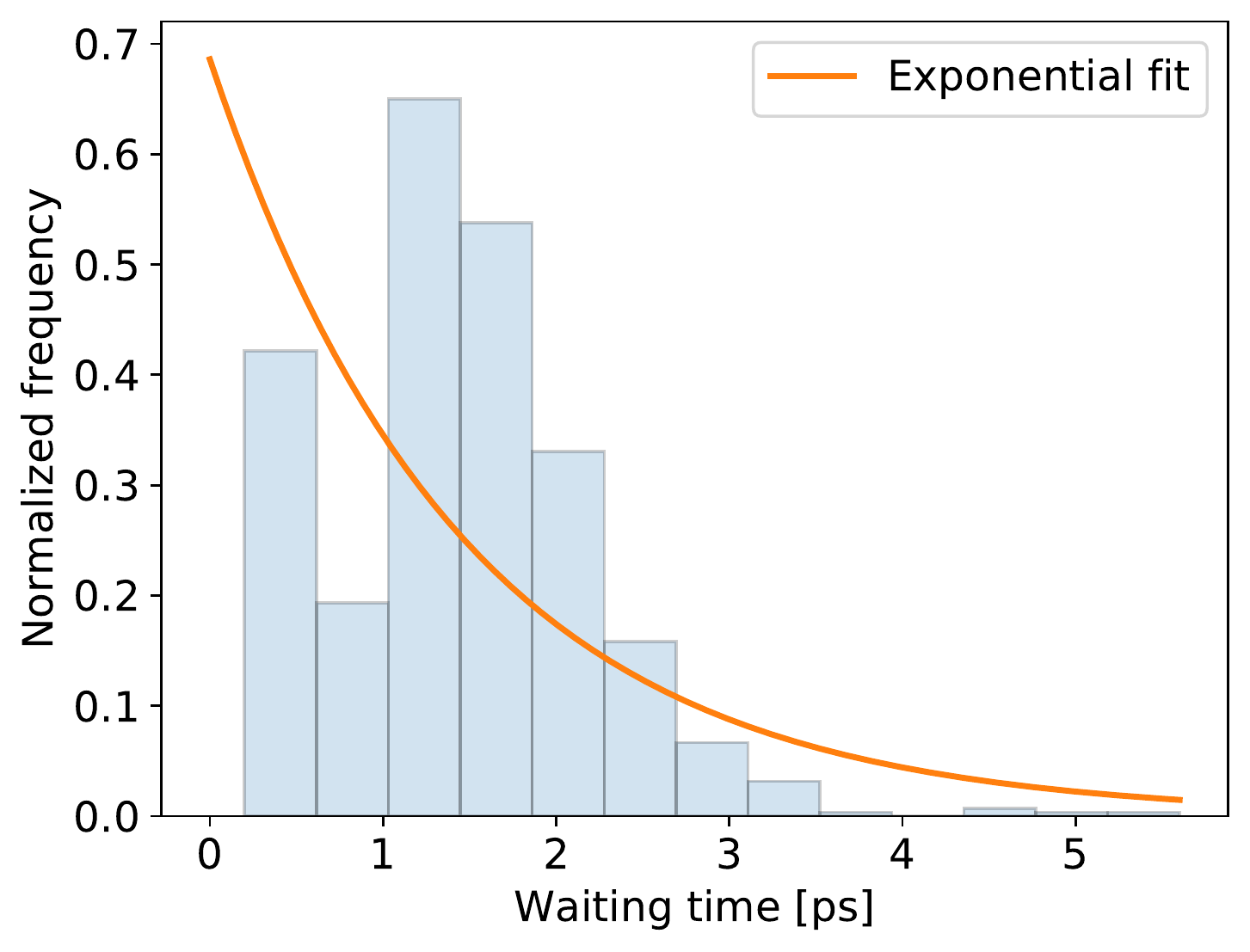}}\\
	\caption{Normalized histograms of waiting times between forward (a) and (d), backward (b) and (e), and any jump (c) and (f), along an exponential fitted curve resulted from MLE parameter estimation for $\tau = 25\ MPa$ (top row) and $\tau = 100\ MPa$ (bottom row).}
	\label{fig:hist}
\end{figure}

Now we show the rate estimation procedure. First, we compile the waiting time statistics in histograms, and plot the normalized histograms with a corresponding exponential fit in Fig.~\ref{fig:hist} for two values of shear stress, $\tau = 25\ MPa$ and $\tau = 100\ MPa$. Observe that distributions of waiting times can be approximated by an exponential decay through its mean value, given the assumption made in the random walk construction.  

We also point that for the lower stress (top row), the distribution of backward waiting times, Fig.~\ref{fig:hist} (c), is closer to the forward case, when compared to a higher stress level (bottom row), Fig.~\ref{fig:hist} (f), which is a direct translation of physical effects that occur at the atomic level into a statistical description of dislocation motion. Furthermore, waiting times for backward jumps at $\tau = 100\ MPa$ are longer than at $\tau = 25\ MPa$, since higher stresses hinder the backward dislocation motion.

From the statistical description of waiting times, we compute the rate constants for forward, backward, and total jumps using the expectation of number of events, Eq.(\ref{eq:rate}), average waiting times, Eq.(\ref{eq:tau}), and the parameter of the exponential fit in Fig.~\ref{fig:hist}, obtained by MLE. Again, we compare results for $\tau = 25\ MPa$ and $\tau = 100\ MPa$, and construct Table~\ref{tab:rates}.

Table~\ref{tab:rates} shows the estimates of $q_f$, $q_b$, and $Q$ directly. We also compute the quantity $q_f + q_b$ and compare it with total rate $Q$ through a relative error measure. We assume that $Q$ is the reference value since it comes directly from data. We observe that all methods yield nearly identical results, specially for $q_f$, which has more available data points. For $q_b$, difference is greater in the $\tau = 100\ MPa$ case due to lower number of data points. We also observe greater error between $q_f + q_b$ and $Q$ for $\tau = 100\ MPs$, for the same reason. 

Nevertheless, the three methods are equivalent, and the differences between their results are negligible, so the choice of any particular method yields nearly identical results in the stochastic simulation. The MLE estimate and the $1/\mu$ result are identical, as expected for the exponential fit. The sample mean estimation from $1/\mu$ should converge to the first case, $\mathbb{E}[N(t)]/t$ as $t\to \infty$ or as $N\to \infty$, since the computation of $\mu$ involves the summation of waiting times, which will approach the total simulation time when the $t$ or $N$ are large. For simplification purposes, for the remaining simulations we will use the expectation estimate, Eq.(\ref{eq:rate}) only due to the agreement between $q_f + q_b$, and total rate $Q$ obtained directly from data points.

\begin{table}[t]
\caption{Rate estimates from MD data for different values of shear stress, using Eq.(\ref{eq:rate}), Eq.(\ref{eq:tau}) and MLE fit.}
\label{tab:rates}
\centering
\begin{tabular}{l|lll|lll}
\hline $\tau$    & \multicolumn{3}{c|}{$25\ MPa$} & \multicolumn{3}{c}{$100\ MPa$} \\ \hline
Method &
  \multicolumn{1}{c}{$\mathbb{E}[N(t)]/t$} &
  \multicolumn{1}{c}{$1/\mu$} &
  \multicolumn{1}{c|}{MLE} &
  \multicolumn{1}{c}{$\mathbb{E}[N(t)]/t$} &
  \multicolumn{1}{c}{$1/\mu$} &
  \multicolumn{1}{c}{MLE} \\ \hline
$q_f$     & 0.633    & 0.634    & 0.634   & 0.625    & 0.626    & 0.626    \\
$q_b$     & 0.352    & 0.353    & 0.353   & 0.060    & 0.062    & 0.062    \\
Q         & 0.985    & 0.987    & 0.987   & 0.685    & 0.686    & 0.686    \\
$q_f+q_b$ & 0.985    & 0.987    & 0.987   & 0.685    & 0.688    & 0.688    \\
Error (\%)    & 0.00    & 0.00    & 0.00   & 0.00    & 0.29    & 0.29     
\end{tabular}
\end{table}

\begin{figure}[t]
	\centering
	\subfloat[$\tau = 25\ MPa$.]{\includegraphics[width=0.33\textwidth]{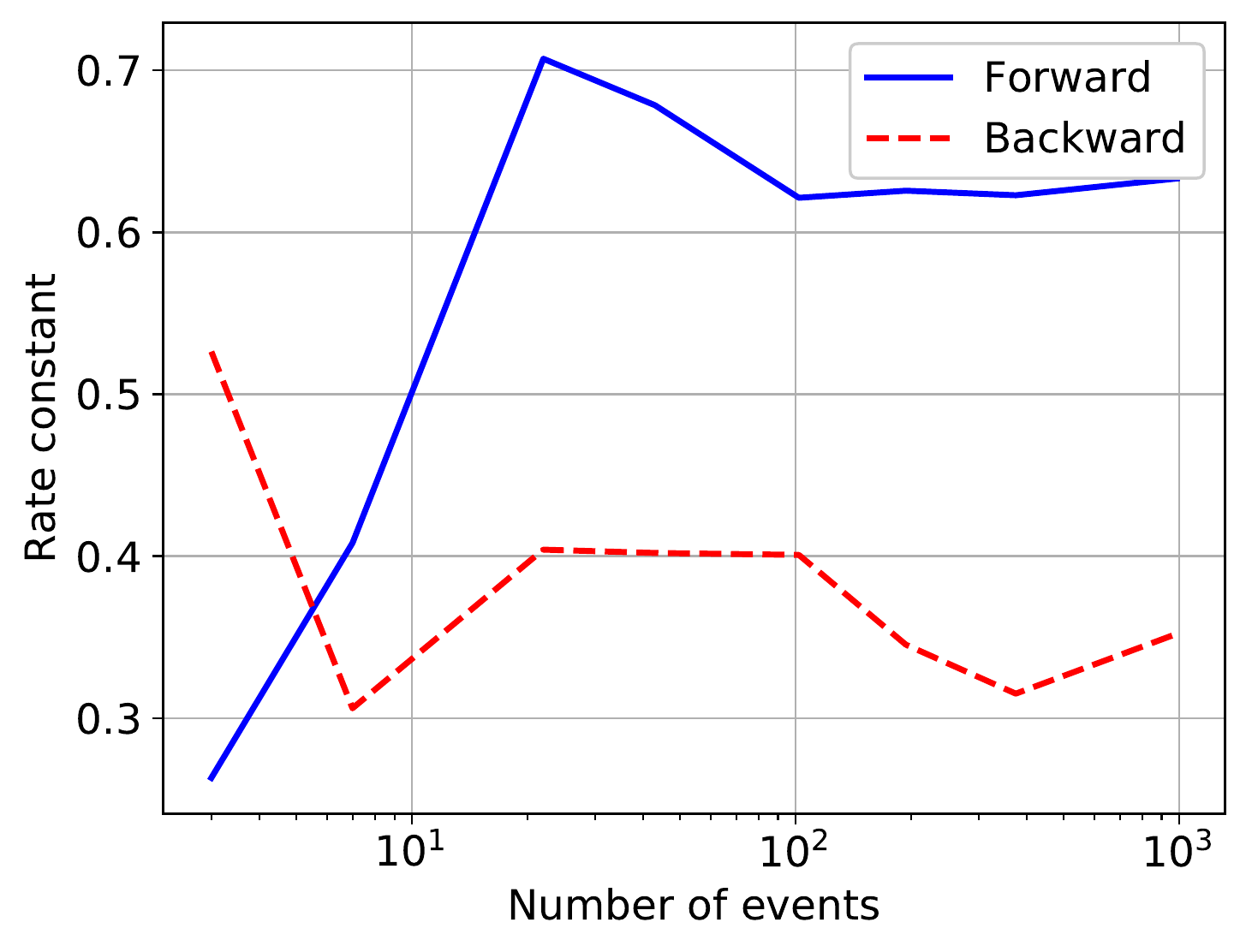}}
	\subfloat[$\tau = 50\ MPa$.]{\includegraphics[width=0.33\textwidth]{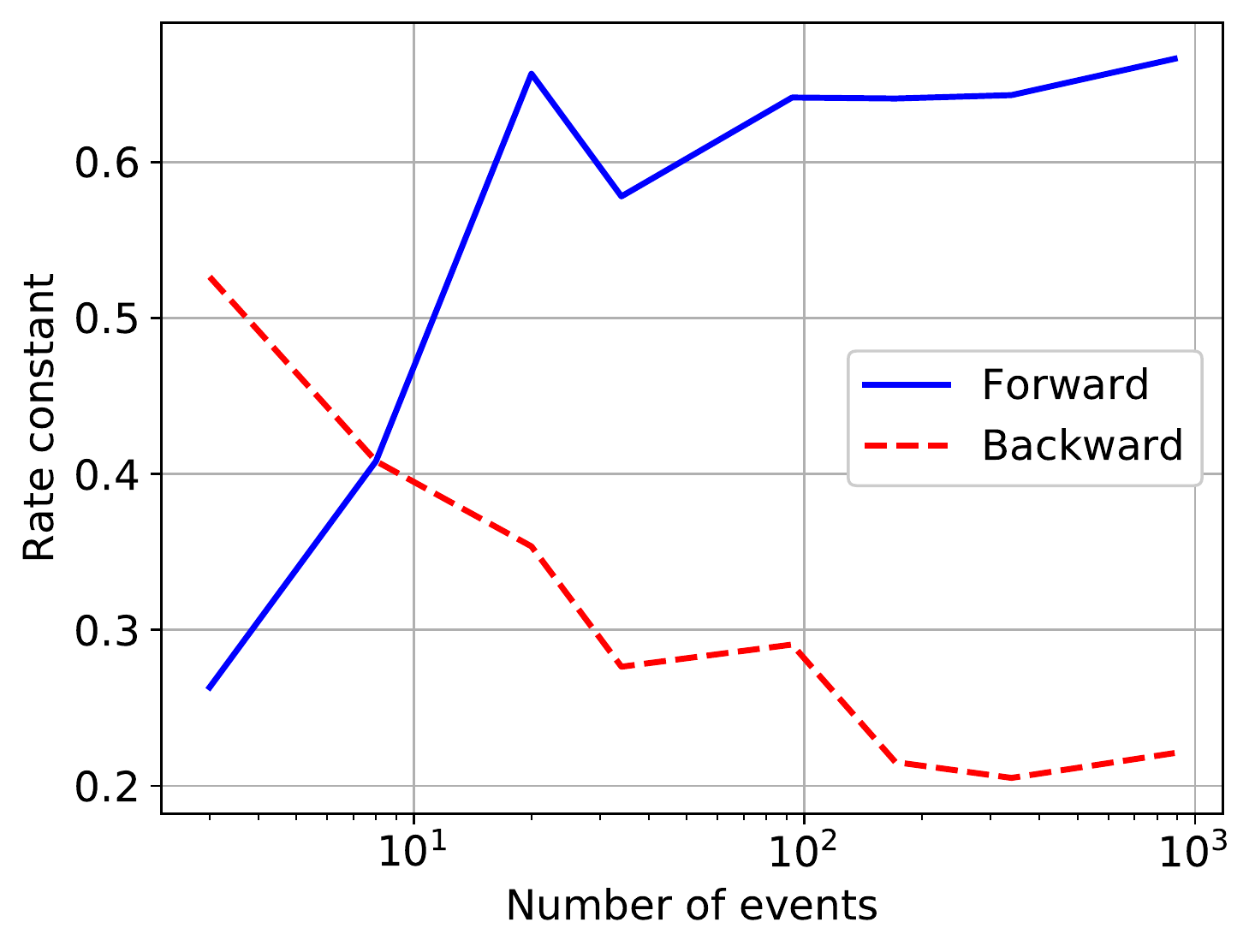}}
	\subfloat[$\tau = 100\ MPa$.]{\includegraphics[width=0.33\textwidth]{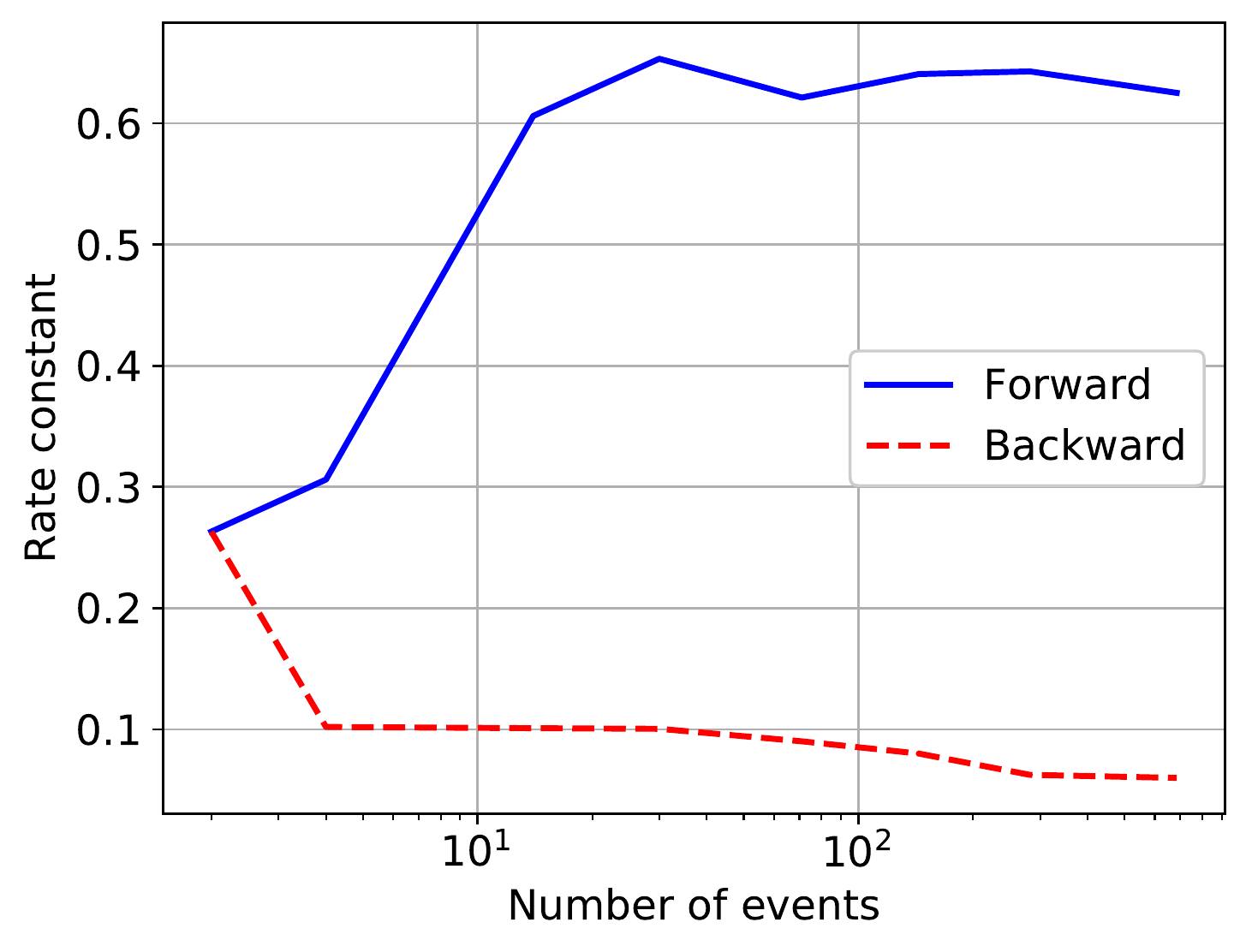}}
	\caption{Convergence in the jump rates from MD time-series data for different stress levels. We observe a more steady and monotonic trend with higher stress levels. }
	\label{fig:md_conv}
\end{figure}

We also check the convergence of estimated rates as in the example with manufactured true rates. Here, we do not know the exact rates, therefore we observe the trend of forward and backward rates as we increase the number of observations. Similarly to the manufactured case, each data point in the plot is generated by considering a truncated time-series, until the final data point which includes the whole time series. Fig.~\ref{fig:md_conv} shows the results of rate estimation, where the $x$-axis again shows the number of statistical events (total number of jumps $N_t(t)$). We observe that the higher the stress level, the smoother is the curve, which is physically consistent. Higher stresses make the forward rates much larger than the backward rates, and the dislocation movement in the MD simulation flows with less noise, so the rate estimates will tend towards a final value with less oscillations.

\subsubsection{Surrogate results}

\begin{figure}[t]
	\centering
	\subfloat[$\tau = 25\ MPa$.]{\includegraphics[width=0.33\textwidth]{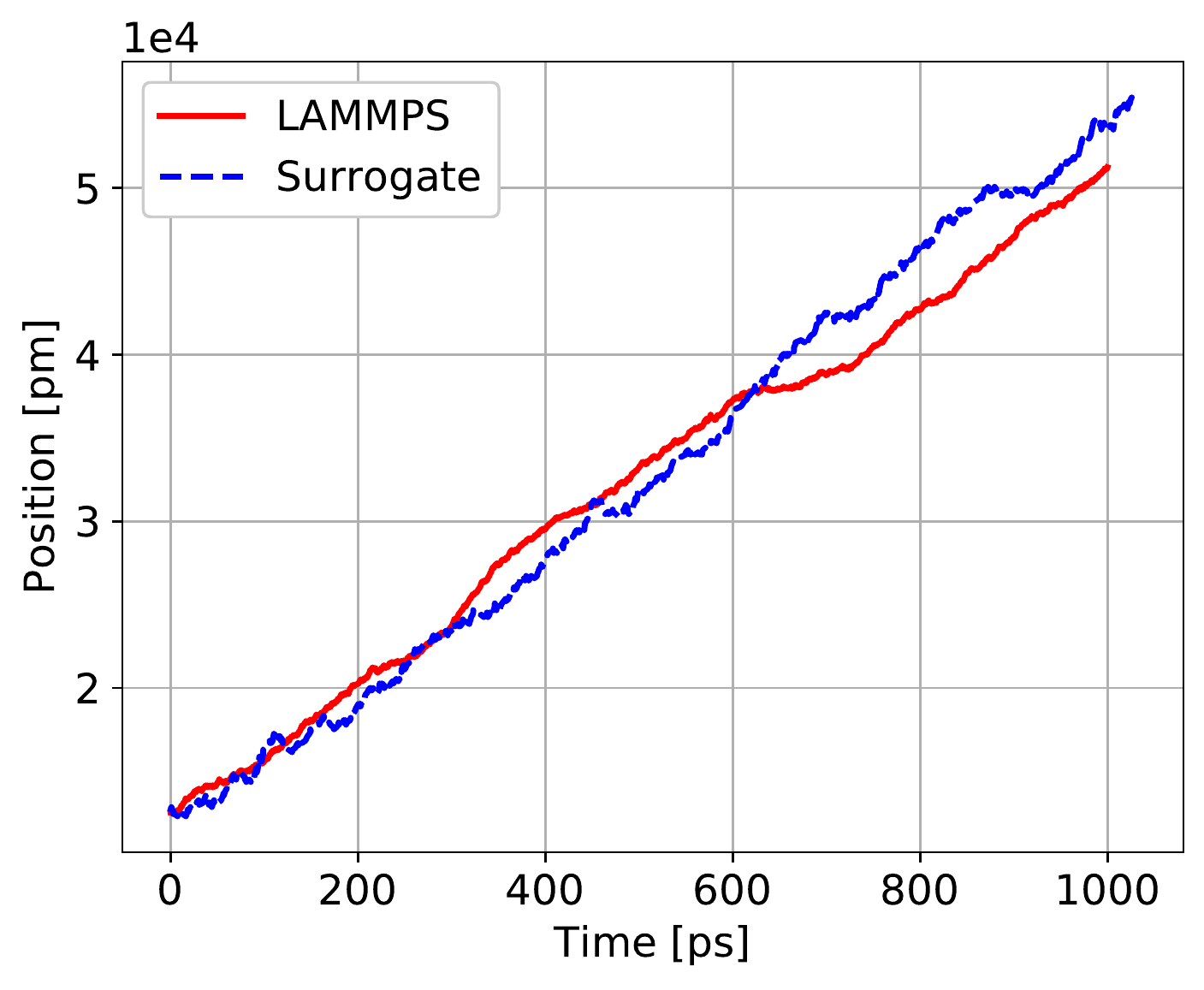}}
	\subfloat[$\tau = 50\ MPa$.]{\includegraphics[width=0.33\textwidth]{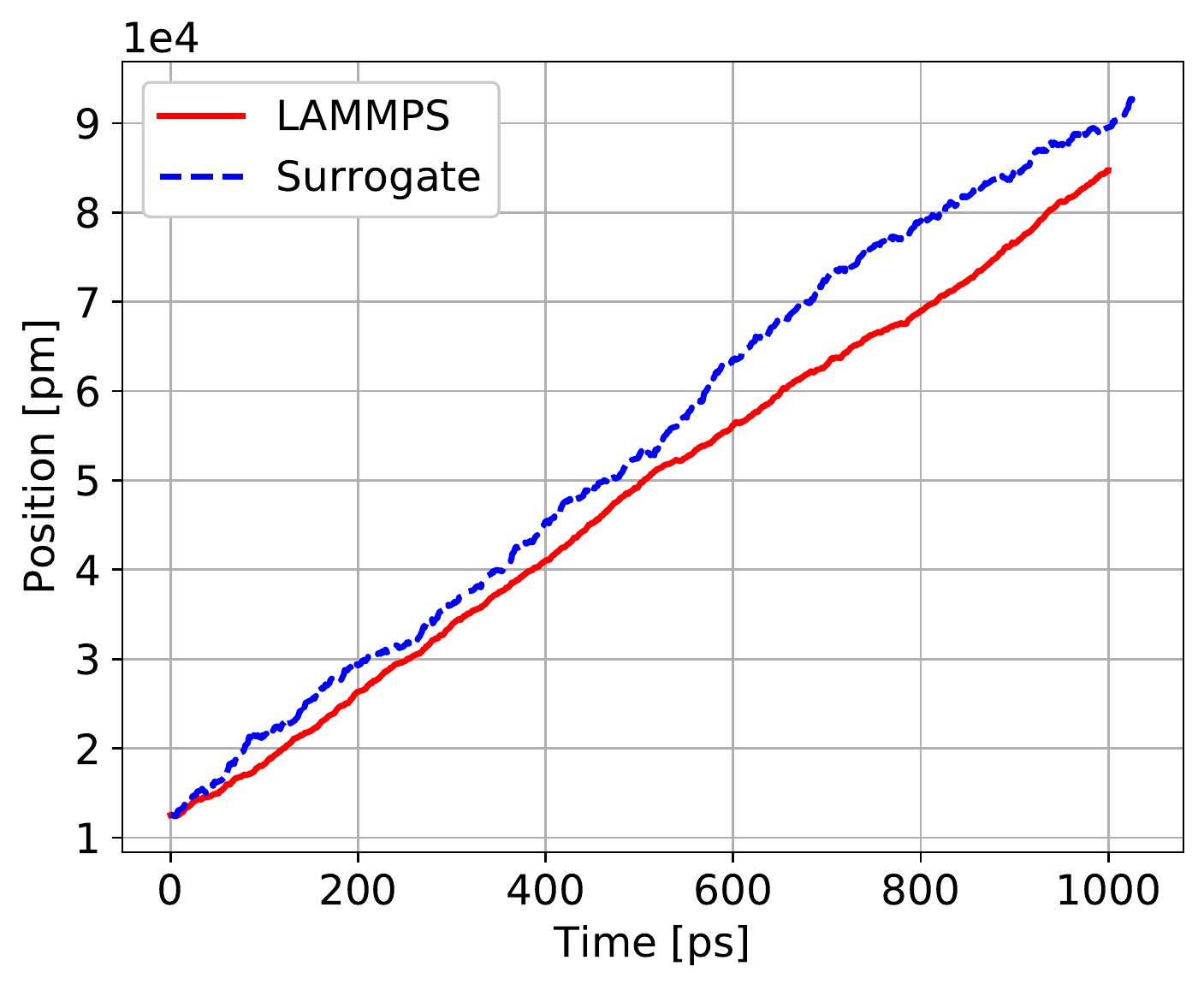}}
	\subfloat[$\tau = 100\ MPa$.]{\includegraphics[width=0.33\textwidth]{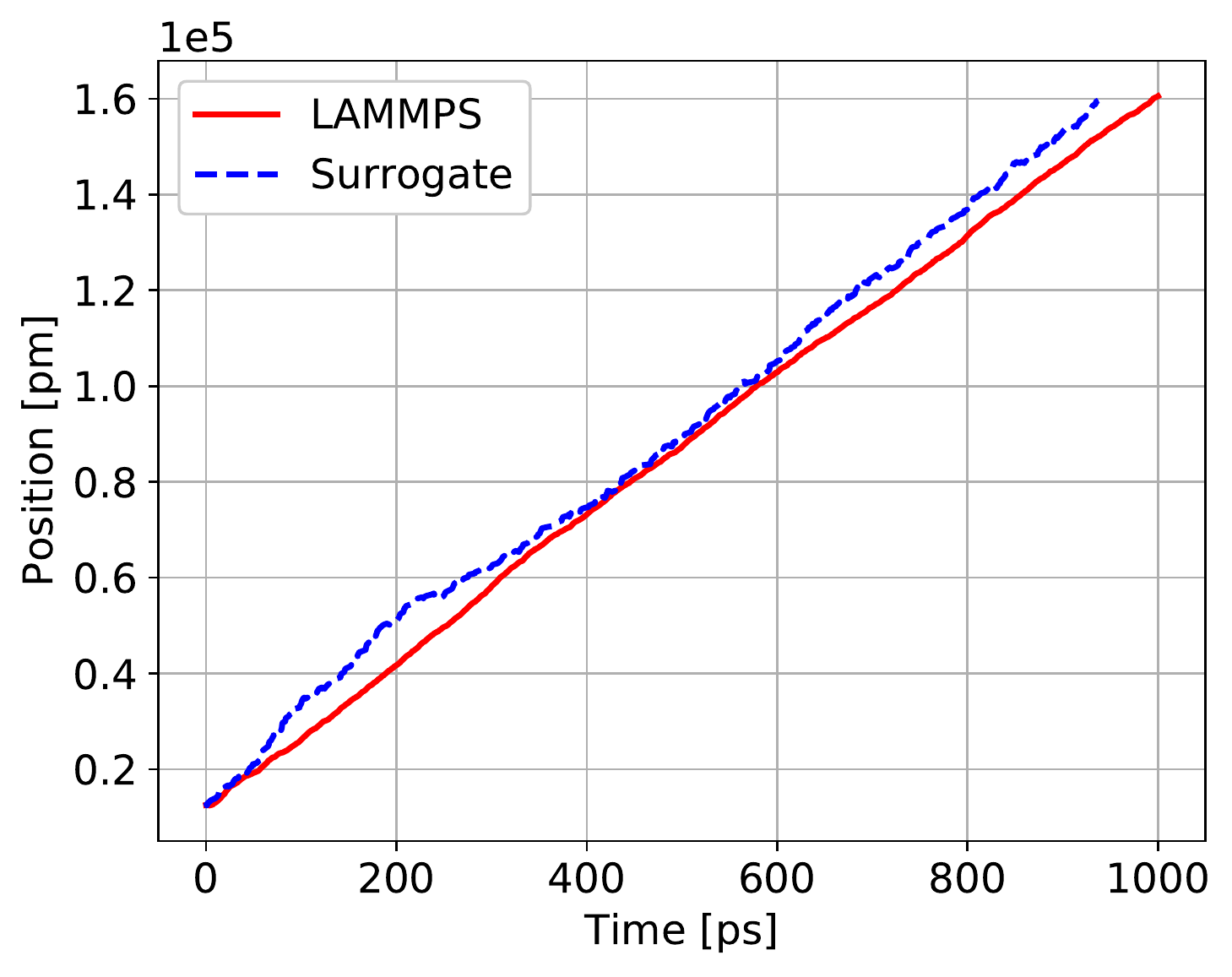}}
	\caption{Position versus time of edge dislocation, comparison between MD results from LAMMPS and one realization of surrogate model through the random walk on a network.}
	\label{fig:position_time}
\end{figure}

For each value of shear stress in the surrogate simulation, we obtain the corresponding rate constants through Eq.(\ref{eq:rate}) and simulate the random walk on a ring graph through the KMC framework, Algorithm~\ref{algo:kmc}. In the end, we are able to plot the distance traveled by the dislocation as a function of time, similar to what is done in MD, by updating the spatial position using Eq.(\ref{eq:jump_update}). We plot the position-time evolution of one realization of the random walk under three different shear stresses, in comparison with the MD results in Fig.~\ref{fig:position_time}.

From Fig.~\ref{fig:position_time} we make some observations. First, under lower stress, MD results are intrinsically noisy, with the dislocation moving more easily under higher stresses, where the MD plot becomes smoother. Those characteristics are manifested in the rate constants as discussed in Table~\ref{tab:rates} and Fig.~\ref{fig:md_conv}, and in the position \textit{versus} time plots generated from the stochastic process in Fig.~\ref{fig:position_time}. 

We also verify form Fig.~\ref{fig:position_time} that the position evolution of the random walk closely follows the same trend as in the original data set. We then compute the dislocation velocity by applying a linear regression model to the plots and computing the slope of the linear fit. We repeat this procedure for a large number of realizations, and run a UQ analysis to obtain the statistics of dislocation mobility.

\begin{figure}[t]
	\centering
	\subfloat[$\tau = 25\ MPa$.]{\includegraphics[width=0.33\textwidth]{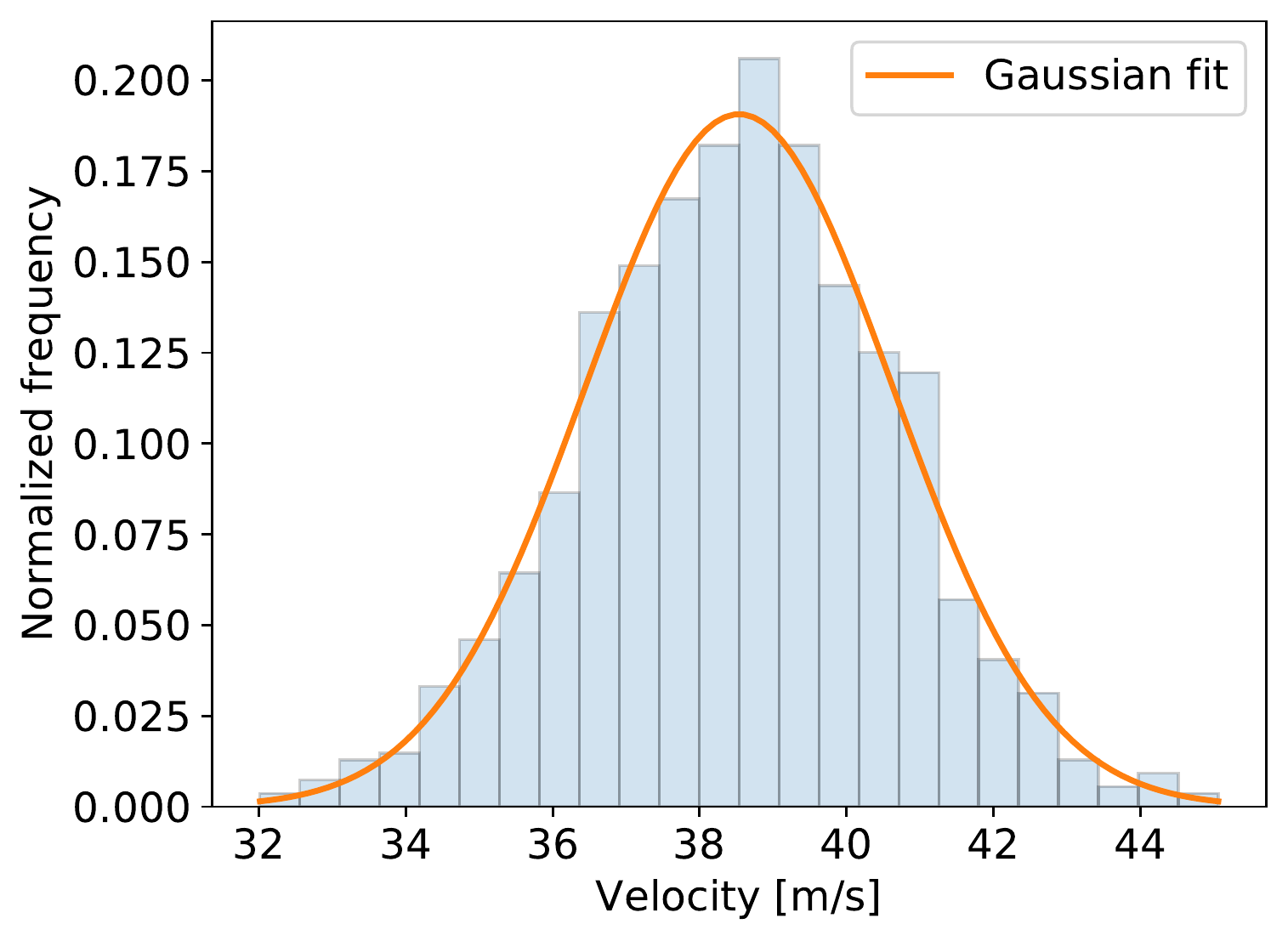}}
	\subfloat[$\tau = 50\ MPa$.]{\includegraphics[width=0.33\textwidth]{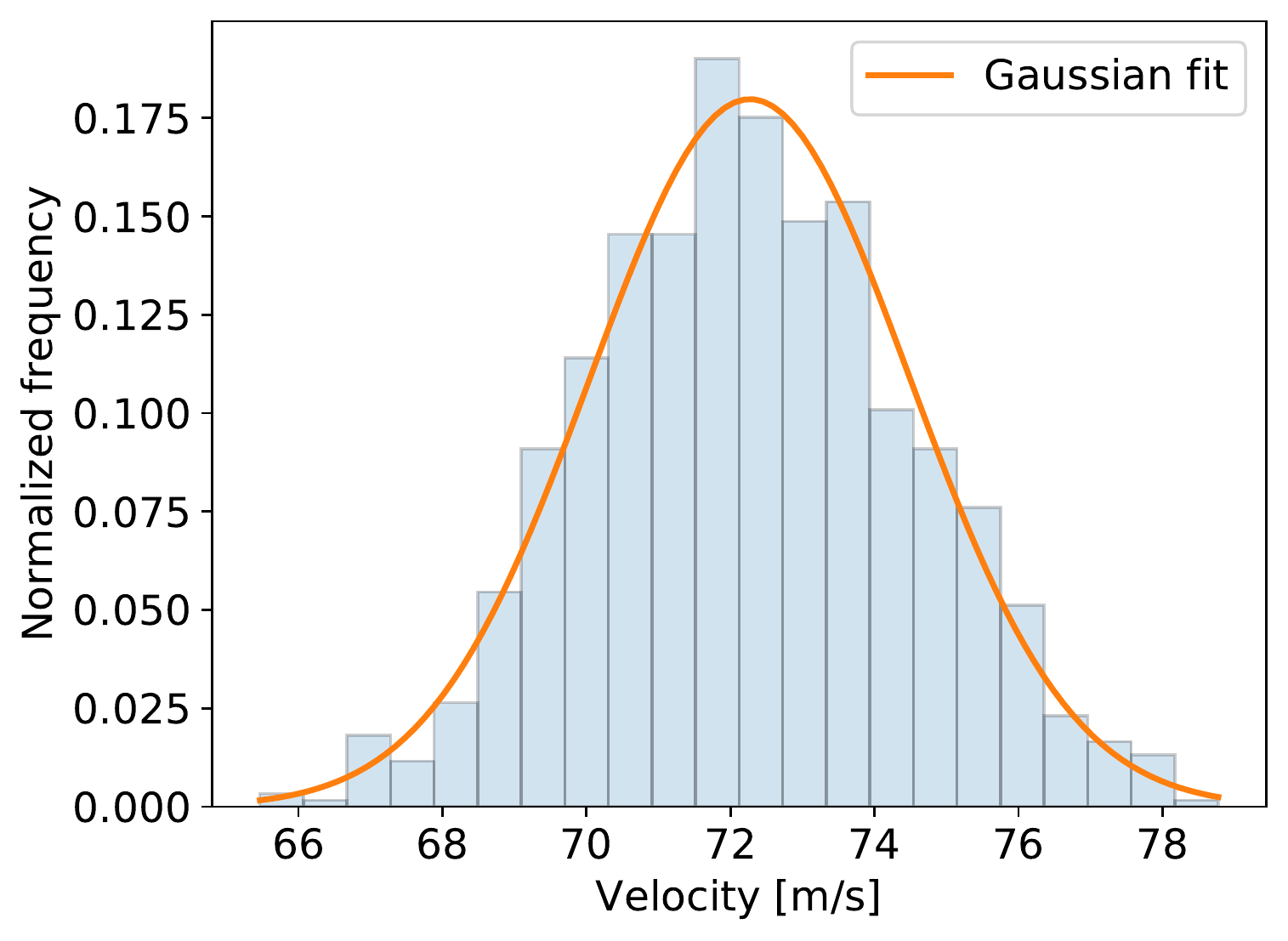}}
	\subfloat[$\tau = 100\ MPa$.]{\includegraphics[width=0.33\textwidth]{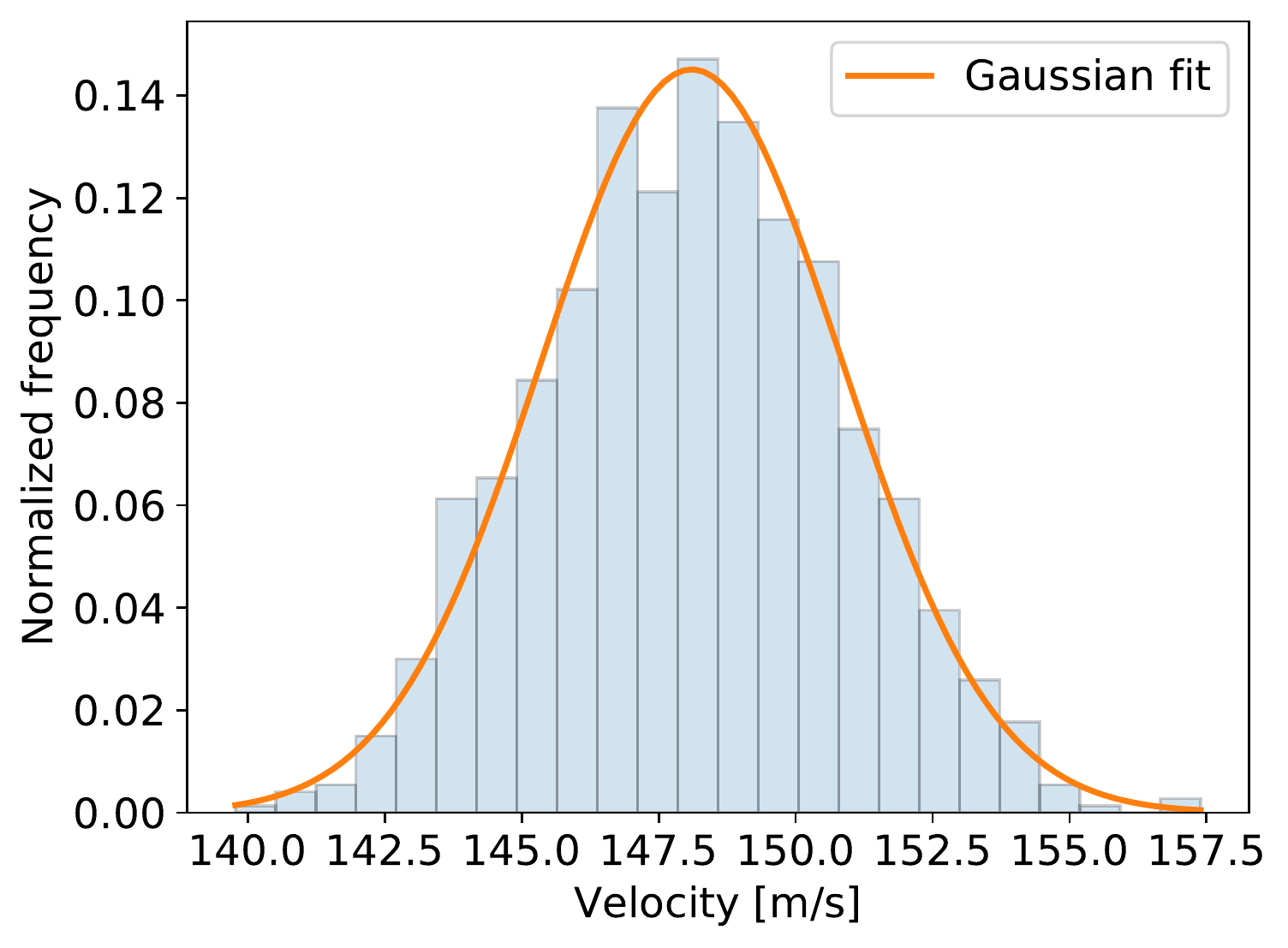}}
	\caption{Normalized histograms of velocity estimates from different applied shear stresses. Gaussian fit is plotted after computation of expectation $\mathbb{E}\left[ v\right]$ and standard deviation $\sigma^2\left[ v\right]$ from 1000 MC realizations.}
	\label{fig:hist_vel}
\end{figure}

We use a simple MC framework to run several realizations of the surrogate simulation, and we obtain the expectation $\mathbb{E}\left[ v\right]$, and standard deviation $\sigma^2\left[ v\right]$ of dislocation velocity under each value of stress. We collect velocity results under $\tau = 25\ MPa$, $\tau = 50\ MPa$, and $\tau = 100\ MPa$, and plot the histograms in Fig.~\ref{fig:hist_vel}. Using the estimated values of $\mathbb{E}\left[ v\right]$ and $\sigma^2\left[ v\right]$ we approximate a Gaussian to the velocity distributions, closely following the histogram. The agreement between the curve and the histogram comes from the Central Limit Theorem \cite{meerschaert2011stochastic}, given that the total simulation time of the surrogate is a summation of exponentially distributed random variables $X$.

\begin{figure}
	\centering
	\includegraphics[width=0.65\textwidth]{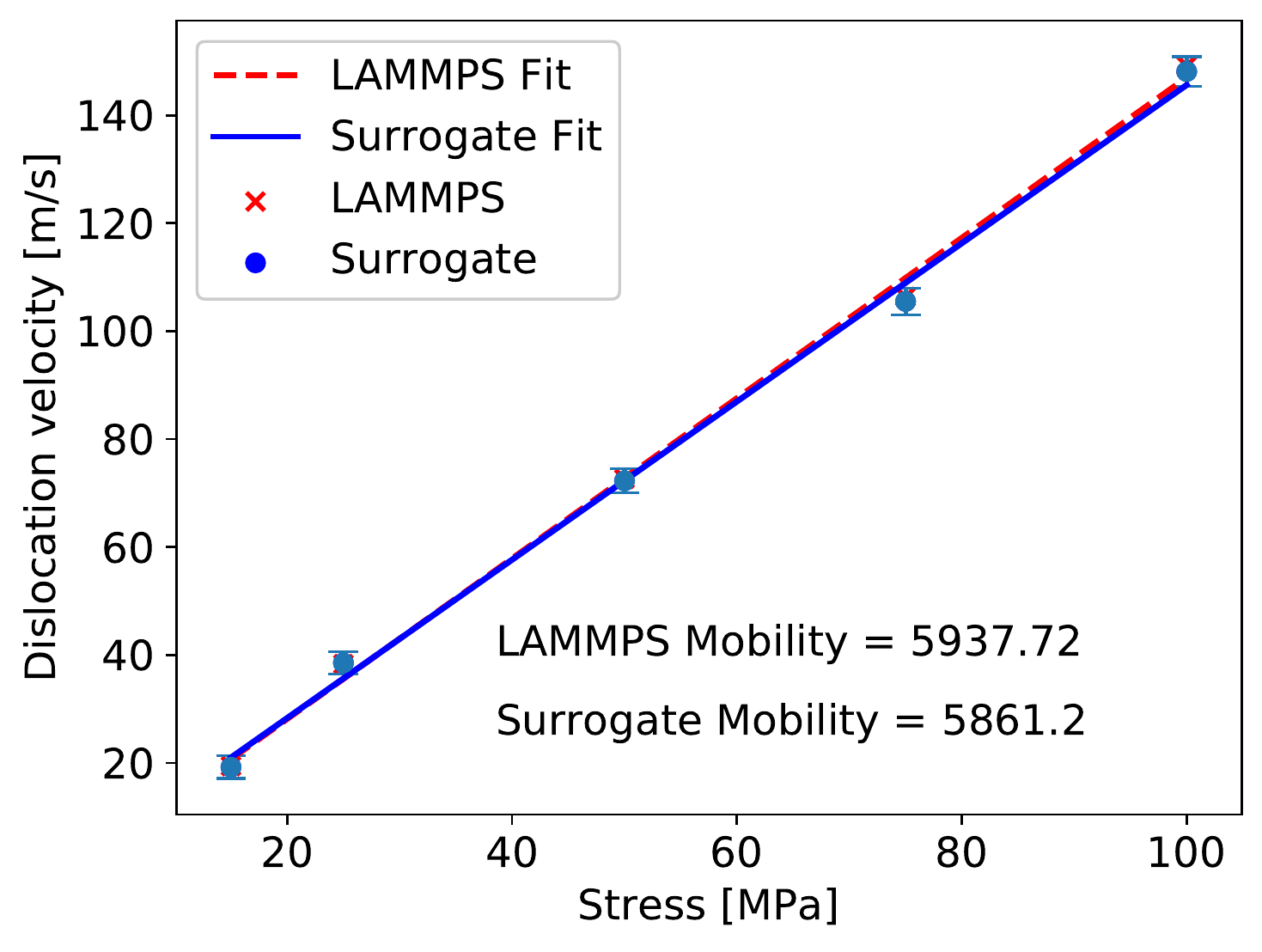}
	\caption{Velocity versus stress plot, comparison between MD results of dislocation glide from LAMMPS, and surrogate model simulations using a random walk in a network under two different system temperatures. The surrogate model accurately estimates the mobility with $1.29\%$ relative error.}
	\label{fig:mob}
\end{figure}

We plot the results of velocity as a function of applied stress in Fig.~\ref{fig:mob}, where we show the expected velocity value, and its corresponding uncertainty represented as error bars, for 1000 MC realizations of the surrogate model. We apply a linear regression model to the velocity-stress plot and obtain the mobility $M$ using the linear fit slope $m$, as in Eq.(\ref{eq:mobility_rule}).

%\begin{figure}
%	\centering
%	\subfloat[$T = 300\ K$]{\includegraphics[width=0.5\textwidth]{mob_300K}}
%	\subfloat[$T = 750\ K$]{\includegraphics[width=0.5\textwidth]{mob_750K}}
%	\caption{Velocity versus stress plot, comparison between MD results of dislocation glide from LAMMPS, and surrogate model simulations using a random walk in a network under two different system temperatures. The surrogate model accurately estimates the mobility with $1.60\%$ and $0.77\%$ relative error for $300$ and $750\ K$, respectively.}
%	\label{fig:mob}
%\end{figure}

By introducing the expected velocity with corresponding uncertainty, as in Fig.~\ref{fig:mob}, we can propagate the uncertainty to the computation of mobility itself. For the set of 1000 realizations shown in Fig.~\ref{fig:mob}, we obtain the corresponding standard deviation for mobility $\sigma_M = 137.27 \ [1/(Pa.s)]$. This is an important contribution of this framework, as it allows a multi-scale propagation of uncertainties related to material properties, starting with the mobility estimate through its modeling as a Poisson process. 

\subsection{Discussion}

Through the definition of a KMC algorithm for a random walk defined on a ring graph topology, where the jump rates are computed directly from time-series data of dislocation motion from an MD simulation, we successfully reproduced the stochastic motion of a dislocation glide in a bcc crystal. The computational advantage of this procedure is two-fold. First, the coarse-graining lumps all the atomic domain information into the network topology, with the dislocation represented as a random walker. The atomistic degrees-of-freedom are condensed into the $n$ nodes that define the graph. Second, we are able to reach the same simulation time faster, which allows for longer time integration, due to the computation of waiting time statistics that feed the KMC algorithm. In the end, 94 hours of one MD simulation with postprocessing at a single stress level turns into an average of 0.45 second surrogate simulation. If we consider the MC estimation of the mobility with 1000 runs at each stress level, the surrogate takes around 50 minutes. 

One important aspect is that the physics of dislocation motion is embedded in the time-series data originated from the MD simulation. Therefore, the computation of process rates of forward and backward jumps already takes that into account from the data itself. This is evident, for example, in Fig.~\ref{fig:md_conv}, where the effect of higher stress rates applied to the atomistic structure translates into higher forward jump rates and lower backward rates. Much of the physics of dislocation motion is embedded in the jump rates, and it would be natural to expand this reasoning to other physical features beyond stress. The characterization of process rates in this broader parametric space can then be achieved with the use of state-of-the-art machine learning (ML) algorithms, with MD simulations used as training data, for a more effective and robust upscaling of dislocation properties.

Furthermore, the mobility uncertainty can be propagated to higher scales to be used as an input with associated error, \textit{e.g.}, in DDD simulations. Later, outputs from  stochastic DDD may be used to inform lumped-element models of elasto-visco-plasticity, or even phase-field models of failure. Through the use of this surrogate model, we provide a quick and efficient method for propagation of uncertainties across scales, starting form the uncertainty estimation at the atomistic level.

\section{Summary and Conclusions}
\label{sec:conclusion}

We developed a data-driven framework for constructing a surrogate model of dislocation glide. Atomistic simulations of dislocation motion provide the statistics that inform the underlying stochastic process of the surrogate. This is achieved firstly through the coarse-graining of MD domain using a graph-theoretical representation. Over this network, the dislocation is idealized as a random walker jumping between the nodes, where the waiting time distribution is parameterized directly from time-series data obtained in MD simulation. The random walk over the network is simulated through a KMC algorithm based on the waiting times obtained empirically. By tracking the dislocation position over we computed the dislocation velocity for each applied shear stress, which in turn leads to the estimation of dislocation mobility.

We highlight the following observations from the model and its numerical results:

\begin{itemize}
    \item The construction followed the assumption of a memoryless, Markovian process governing the dislocation motion, which was a sufficient description based on an estimate of average waiting times from empirical data.
    \item The estimation of rate constants, often a major difficulty in the application of KMC, was performed directly through MD data. We compared three different methods that yielded nearly identical results.
    \item From the computed rates, the actual simulation of the stochastic process resulted in dislocation motion in agreement with trajectories simulated by MD. Next, computation of mobility through the surrogate also had excellent agreement with original atomistic estimates. 
    \item Simulation through the surrogate achieved remarkable speedup compared with MD computation times.
    \item Uncertainty levels dependent on the number of data points used to construct the surrogate. We provided uncertainty estimates for the mobility through the surrogate, taking into account the variance of the underlying stochastic process.
\end{itemize}

The current framework is still limited in the sense that it only simulates a single dislocation under glide, and it disregards more complex mechanisms that are not consistent with Markovian processes, such as heavy-tailed processes, which may appear during failure \cite{miguel2001intermittent,bonamy2008crackling}. However, multiple dislocations could be simulated by considering additional random walkers in the surrogate model. Furthermore, the construction of the model is still dependent on performing high-fidelity atomistic simulations to obtain the rate constants. However, we provide the groundwork that allows the incorporation of more elaborate physics, with the advantage of running the simulation for longer times due to speedup. 

We emphasize that our proposed framework establishes a meaningful bridge for coupling scales, where not only the value of mobility is provided, but its associated uncertainty. Through the description of dislocation motion as a stochastic process informed by high-fidelity data, we can propagate the uncertainty associated with mobility estimations or any other quantity of interest, even with a limited number of MD samples. As a consequence, this framework acts as a tool for more predictive multi-scale material characterization.

\bibliographystyle{siamplain}
\bibliography{reference}

\end{document}

%% file: shared.tex
\usepackage{amsfonts}
\usepackage{amsmath}
\usepackage{graphicx}
\usepackage{epstopdf}
\usepackage{algorithmic}
\usepackage{mathtools}
\usepackage{bm}
\usepackage[caption=false]{subfig}
\usepackage{float}

\ifpdf
  \DeclareGraphicsExtensions{.eps,.pdf,.png,.jpg}
\else
  \DeclareGraphicsExtensions{.eps}
\fi

\numberwithin{theorem}{section}

\newcommand{\TheTitle}{Atomistic-to-Meso Multi-Scale Data-Driven Graph Surrogate Modeling of Dislocation Glide} 
\newcommand{\TheAuthors}{E. A. Barros de Moraes, J. L. Suzuki and M. Zayernouri}

\headers{}{}

\author{
  Eduardo A. Barros de Moraes\thanks{Department of Mechanical Engineering \& Department of Computational Mathematics, Science and Engineering, Michigan State University, 428 S Shaw Ln, East Lansing, MI 48824, USA.}
  \and
  Jorge L. Suzuki\thanks{Department of Mechanical Engineering \& Department of Computational Mathematics, Science and Engineering, Michigan State University, 428 S Shaw Ln, East Lansing, MI 48824, USA.}
  \and
  Mohsen Zayernouri \thanks{Department of Mechanical Engineering \& Department of Statistics and Probability, Michigan State University, 428 S Shaw Ln, East Lansing, MI 48824, USA, Corresponding Author; \email{zayern@msu.edu}}
}

\title{{\TheTitle}\thanks{This work was supported by the ARO Young Investigator Program Award (W911NF-19-1-0444), the MURI/ARO grant (W911NF-15-1-0562), and partially by the National Science Foundation Award (DMS-1923201). The HPC resources and services were provided by the Institute for Cyber-Enabled Research (ICER) at Michigan State University.}}

\usepackage{amsopn}

\newtheorem{remark}{Remark}